\documentclass[aps,prd,twocolumn,showpacs,amsmath,amssymb,nofootinbib,eqsecnum, preprintnumbers]{revtex4-1}

\usepackage{graphicx}
\RequirePackage{ulem}
\usepackage{amsbsy}
\usepackage{amsmath}
\usepackage{amsthm}

\newcommand{\eps}{\varepsilon}

\newcommand{\pa}{{\partial}}

\newcommand{\be}{\begin{equation}}
\newcommand{\ee}{\end{equation}}
\newcommand{\ba}{\begin{eqnarray}}
\newcommand{\ea}{\end{eqnarray}}

\newcommand{\beq}{\begin{equation}}
\newcommand{\eeq}{\end{equation}}
\newcommand{\beqa}{\begin{eqnarray}}
\newcommand{\eeqa}{\end{eqnarray}}
\newcommand{\nn}{\nonumber}

\newcommand{\bgc}[2]{{\bigl[\mspace{-4.4mu}\lvert#1,#2\rvert\mspace{-4.4mu}\bigr]}}

\newcommand{\hook}{\raisebox{-0.35ex}{\makebox[0.6em][r]
{\scriptsize $-$}}\hspace{-0.15em}\raisebox{0.25ex}{\makebox[0.4em][l]{\tiny
 $|$}}}

\newcommand{\cwedge}[1]{\mathop{\wedge}_{{}^{#1}} }

\newcommand{\dm}{n}

\newcommand{\M}{\mathcal{M}}
\newcommand{\oM}{\hat{\mathcal{M}}}
\newcommand{\og}{\hat{g}}

\newcommand{\oGamma}{\hat{\Gamma}}
\newcommand{\oX}{\hat{X}}
\newcommand{\oomega}{\hat{\omega}}
\newcommand{\of}{\hat{f}}
\newcommand{\onabla}{\hat{\nabla}}
\newcommand{\oE}{\hat{E}}
\newcommand{\eh}{\hat{e}}
\newcommand{\h}[1]{\hat{#1}}

\makeatletter
\newcommand{\notes}{%
  \clearpage
  \section*{Notes}%
  \addcontentsline{toc}{section}{Notes \hrulefill}
  \setcounter{section}{0}%
  \setcounter{subsection}{0}%
  \renewcommand\thesection{\@Roman\c@section}%
  }
\makeatother

\newcommand\fverb{\setbox\fverbbox=\hbox\bgroup\verb}
\newcommand\fverbdo{\egroup\medskip\noindent%
			\fbox{\unhbox\fverbbox}\ }
\newcommand\fverbit{\egroup\item[\fbox{\unhbox\fverbbox}]}
\newbox\fverbbox

\begin{document}

\title{Hidden symmetries of Eisenhart-Duval lift metrics and the Dirac equation with flux}

\author{Marco Cariglia}
\email{marco@iceb.ufop.br}
\affiliation{Universidade Federal de Ouro Preto, ICEB, Departamento de F\'isica.
  Campus Morro do Cruzeiro, Morro do Cruzeiro, 35400-000 - Ouro Preto, MG - Brasil}

\date{June 21, 2012}

\begin{abstract}
The Eisenhart-Duval lift allows embedding non-relativistic theories into a Lorentzian geometrical setting. In this paper we study the lift from the point of view of the Dirac equation and its hidden symmetries. We show that dimensional reduction of the Dirac equation for the Eisenhart-Duval metric in general gives rise to the non-relativistic L\'evy-Leblond equation in lower dimension. We study in detail in which specific cases the lower dimensional limit is given by the Dirac equation, with scalar and vector flux, and the relation between lift, reduction and the hidden symmetries of the Dirac equation. While there is a precise correspondence in the case of the lower dimensional massive Dirac equation with no flux, we find that for generic fluxes it is not possible to lift or reduce all solutions and hidden symmetries. As a by-product of this analysis we construct new Lorentzian metrics with special tensors by lifting Killing-Yano and Closed Conformal Killing-Yano tensors and describe the general Conformal Killing-Yano tensor of the Eisenhart-Duval lift metrics in terms of lower dimensional forms. Lastly, we show how dimensionally reducing the higher dimensional operators of the massless Dirac equation that are associated to shared hidden symmetries it is possible to recover hidden symmetry operators for the Dirac equation with flux.	
\end{abstract}




\maketitle


\section{Introduction}
\noindent The Eisenhart-Duval lift of a Riemannian metric \cite{Eisenhart1928} is an example of geometrisation of interactions. The dynamics of a classical physical system, described by a Riemannian metric $h_{\mu\nu}$ in $\dm$ dimensions and in the presence of a scalar potential $V$ and a vector potential $A_\mu$, is shown to be equivalent to geodesics in a Lorentzian spacetime of dimension $\dm +2$. This geometrical idea has been historically introduced by Eisenhart in \cite{Eisenhart1928}, however to our knowledge it took a number of years after the same idea was independently re-discovered in \cite{DuvalBurdetKunzlePerrin1985}, from there prompting further work, among which \cite{GaryDuvalHorvathy1991,GaryPeterHorvathyZhang2012}. Over time it has found  a number of applications, among which one can mention the following, non-exhaustive, examples: providing a relativistic framework to study non-relativistic physics, as the free Schr\"{o}dinger equation in $\dm$ dimensions and with metric $g$ can be written in the lifted geometry as the free, massless Klein-Gordon equation \cite{HorvathyZhang2009}; simplifying the study of symmetries of a Hamiltonian system by looking at geodesic Hamiltonians \cite{Benenti1997,Benn2006}; building new Lorentzian pp-wave metrics solutions of the Einstein-Maxwell equations \cite{GaryChrisPope2010,GaryPeterHorvathyZhang2012}; studying from a geometrical point of view dynamical systems as diverse as protein folding \cite{MazzoniCasetti2008}, rare gas crystals \cite{CasettiMacchi1996} and chaotic gravitational $N$-body systems \cite{Cerruti-SolaPettini1995}. 

On a separate account, there has been much recent activity in the study of hidden symmetries of physical systems. The interest has increased for two main reasons. First it has been discovered that a number of hidden symmetries are related to separation of variables for equations of physics related to different spin and of either classical or quantum nature: the geodesic equation  \cite{DavidPavelPageVasudevan2006}, the Hamilton-Jacobi and Klein-Gordon equation \cite{DavidPavelFrolov2007,PavelSergyeyev2008}, the Dirac equation \cite{OotaYasui2007,MDP2011_2}, electromagnetic perturbations in $\dm = 5$ dimensions \cite{MurataSoda2007}, linearised gravitational perturbations \cite{MurataSoda2007,KodamaIshibashi:2003, KunduriEtal:2006, OotaYasui:2010, Dias:2010a}. Such separation has been achieved for Kerr-NUT-(A)dS spacetimes \cite{ChenEtal:2006cqg}, which are  higher-dimensional generalisations of the Kerr metric. For these spacetimes it has been proven that a Principal Conformal Killing-Yano (PCKY) tensor is present \cite{DavidFrolov2006} and that from it all the geometrical structure and further hidden symmetries follow \cite{DavidFrolov2008}. It is also possible to show that the theory of the worldline supersymmetric spinning particle in these spacetimes admits a number of non-trivial supercharges that make its bosonic sector integrable \cite{DavidMarco2011}. 
In general see \cite{DavidThesis,HouriYasui2011,Santillan2011} for an extensive review of hidden symmetries in the framework of gravitational systems. The second reason for the recent activity is the fact that several new examples of spacetimes with non-trivial hidden symmetries of higher order have been found, in many cases using the Eisenhart-Duval lift procedure applied to integrable systems such as the Goryachev-Chaplygin top, the Kovalevskaya top, the Calogero model \cite{DavidGaryetAl2011,GaryRugina2011,Galajinsky2012}. 
 
In this paper we look at the Eisenhart-Duval lift procedure from the point of view of hidden symmetries of the Dirac equation and of Conformal Killing-Yano tensors (CKY). There is a number of reasons why this is meaningful. First, if it is possible to perform the Eisenhart-Duval lift of a known CKY tensor then this opens the possibility to create new Lorentzian metrics with CKY tensors. To this extent it is useful to note that a classification of higher-dimensional spacetimes with CKY tensors has only been completed in the case of rank 2 closed tensors and Riemannian signature with and without torsion \cite{DavidPavelFrolov2008_uniqueness,HouriOotaYasui2009,AhmedovAliev2009,DavidClaudeHouriYasui2012}. In this paper we show how to perform such a lift under appropriate conditions, thus presenting new Lorentzian metrics with CKY tensors. Second, since the Eisenhart-Duval lift links a higher dimensional dynamics in the absence of forces other than gravity to that of a lower dimensional system in the presence of scalar and vector potential, there is the possibility to establish a link between the higher-dimensional (massless) Dirac equation and a Dirac equation in lower dimension with scalar and vector flux. In the paper we show concretely how to perform the dimensional reduction of the higher dimensional Dirac equation and obtain the lower dimension Dirac equation with flux, and its inverse operation, oxidation. This is of interest in itself and more so since we are able to show a geometrical link between hidden symmetry operators of the free higher dimensional Dirac equation, which are given in terms of CKY tensors \cite{BennCharlton:1997, BennKress:2004}, and the recently discussed hidden symmetry operators of the Dirac equation with flux \cite{DavidClaudePavel2011}. While analising the hidden symmetries of the higher dimensional and lower dimensional theories we find the non-trivial result that for generic fluxes each of the two theories can have hidden symmetries that are not present in the other. It is worth noticing that this phenomenon is not present when we relate the massive lower dimensional Dirac equation with the other fluxes turned off and the massless higher dimensional Dirac equation. For those hidden symmetries that are common to the two theories we can perform dimensional reduction and find the symmetry operators with flux discussed in \cite{DavidClaudePavel2011}. Under the hypothesis that symmetry operators linear in momenta for the lower dimensional theory cannot be lifted to symmetry operators in higher dimension of order $>1$, we interpret this result as meaning that the two theories differ as the level of phase space dynamics, which is different from what happens in the case of a scalar particle. A third reason to study CKY tensors for Eisenhart-Duval lift metrics is that it is possible to characterise the higher dimensional CKY equation completely in terms of lower dimensional forms. We do this and obtain equations in lower dimensions that generalise the CKY equation and implicitly classify the most general higher dimensional CKY tensor. 

The rest of the paper is organised as follows. In section \ref{sec:preliminaries} we introduce useful notation and basic notions about hidden symmetries. Section \ref{sec:Eisenhart} 
is devoted to the Eisenhart-Duval lift. We discuss the geometrical lift and the dynamics of a scalar particle and its hidden symmetries. Finally we show how to lift lower dimensional CKY tensors and the fact that there are restrictions to this procedure. We also classify the most general higher dimensional CKY tensor in terms of lower dimensional forms. Section \ref{sec:Dirac} is devoted to analysing the dimensional reduction of the Dirac equation in higher dimension. We show that in general one can recover in lower dimension the non-relativistc L\'evy-Leblond equation. In addition, it is also possible in some cases to recover the lower dimensional Dirac equation with flux using the higher dimensional massless Dirac equation plus a non-trivial projection. Such projection is responsible for the fact that not all higher dimensional linear symmetry operators can be dimensionally reduced: for a specific class of symmetry operators we show in detail when this can be done and obtain a subset of the linear symmetry operators of the Dirac equation with flux discussed in \cite{DavidClaudePavel2011}. We also use the Dirac equation to gain insight on the earlier finding that not all lower dimensional CKY tensors can be lifted. Section \ref{sec:examples} presents some examples and finally section \ref{sec:conclusions} is devoted to conclusions and final remarks. In the appendix we discuss useful identities for the metric, Hodge duality, differentiation of forms and we present the full set of higher dimensional CKY equation in terms of equations for lower dimensional forms.


\section{Preliminaries\label{sec:preliminaries}}

\subsection{Notation} 
We start with a Riemannian metric $g$ defined on an $\dm$--dimensional manifold $\M$. Its Eisenhart-Duval lift will be an $(\dm + 2)$--dimensional manifold $\h{\M}$ which is a bundle over $\M$ on which a Lorentzian metric $\h{g}$ is defined. In general $\dm +2$--dimensional quantities will be denoted with a hat symbol, so for example if $f$ is a $p$--form on $\M$ then its natural embedding in $\h{\M}$ will be denoted by $\h{f}$, if $D$ is the Dirac operator on $\M$ then $\h{D}$ will be the Dirac operator on $\h{\M}$ and so on. Indices $\mu, \nu, \dots$, from the lowercase Greek alphabet represent spacetime indices on $\M$, while $M, N, \dots$, from the uppercase Latin alphabet spacetime indices on $\h{\M}$. Local coordinated used for $\M$ are $\{ x^\mu \}$, and for the lift we introduce new variables $v, t$ so that $\{ \h{x}^M \} = \{ v, t, \{x^\mu \}\}$ are local coordinates on $\h{\M}$. When we work with the Dirac equation and Gamma matrices it is convenient to use  locally flat indices: we will use $a = 1, \dots , \dm$ for $\M$ and $A = +, -, 1, \dots, \dm)$ for $\h{\M}$. Vielbein forms on $\M$ are indicated as $e^a = e^a_\mu dx^\mu$ and, analogously, on $\h{\M}$ as $\h{e}^A = \h{e}^A_M d \h{x}^M$. Inverse vielbeins are written as $E^\mu_a$ and, respectively, $\h{E}^M_A$. $\nabla$ always means the appropriate covariant derivative, acting on either tensors, forms or spinors. 

The notation we use to describe differential forms is the following - we display formulas valid on $\M$, and similar formulas hold for $\h{\M}$. Let $\{ dx^\mu \}$ be a coordinate basis for $1$--forms, and $\{ \partial_\mu \}$ for vectors.  The exterior algebra is ${\Omega(\M)=\bigoplus_{p=0}^{\dm} \Omega^p(\M)}$. Given a $p$--form $ \omega = \frac{1}{p!} \omega_{\mu_1 \dots \mu_p} dx^{\mu_1} \dots dx^{\mu_p} \in \Omega^p(\M)$ and a vector $v$ the inner derivative of $\omega$ relative to $v$, or hook operation, is a $(p-1)$--form $v \hook \omega$ with components given by 
\be 
(v \hook \omega)_{\mu_1 \dots \mu_{p-1}} = v^\lambda \omega_{\lambda \mu_1 \dots \mu_{p-1}} \, . 
\ee 
Given a vector $v = v^\mu \partial_\mu$ there is a canonical form associated to it using the metric to lower the component indices, $v^\flat = v_\mu dx^\mu$, and similarly given a $1$--form $\lambda = \lambda_\mu dx^\mu$ there is a vector $\lambda^\sharp = \lambda^\mu \partial_\mu$. Such operation is also called musical isomorphism. Given a vielbein basis for $1$--forms $\{ e^a \}$  then the vectors $X^a = (e^a)^\sharp$ are a basis for the tangent space of $\M$ and satisfy $(X^a )^\mu (X^b)_\mu = \eta^{ab}$ where $\eta$ is the unit matrix for $\M$ (and $\h{\eta}$ is the Minkowski metric for $\h{\M}$).  Then the differential and co-differential of a form $\omega$ can be written as 
\be 
d \omega = e^a \wedge \nabla_a \omega \, , \qquad \delta \omega = - X^a \hook \nabla_a \omega \, . 
\ee 
An inhomogeneous form $\omega$ can be written as a sum of homogeneous $p$-forms
\begin{equation}\label{inhomforms}
\omega = \sum_{p=0} \omega^{(p)} \; . 
\end{equation}
We define  the degree operator ${\pi}$ and parity operator ${\eta}$ which act as
\begin{equation}\label{pietadef}
\pi \omega = \sum_{p=0} p \, \omega^{(p)} \;,\quad
\eta \omega = \sum_{p=0} (-1)^p \omega^{(p)} \, .
\end{equation}

For $\alpha$, $\beta$ a $p$-- and, respectively, $q$--form, we define the {\it contracted wedge product} recursively by 
\ba
\alpha\cwedge{0}\beta &=& \alpha \wedge \beta  \, , \nn \\
\alpha\cwedge{k}\beta &=& (X_a \hook \alpha ) \cwedge{k-1} (X^a \hook \beta) \qquad (k\ge 1) \, , \nn \\
\alpha\cwedge{k}\beta &=& 0 \qquad\qquad\qquad\qquad\qquad\, (k<0) \, .\label{cwedgedef}
\ea
The contracted wedge product satisfies the identities 
\begin{equation}\label{mywedge_hookWedge}
\begin{aligned}
e^a \wedge \left[ \left( X_a \hook \alpha \right) \cwedge{m} \beta \right] &=
  (-1)^m\, \left[ \left( \pi-m \right)\alpha \right] \cwedge{m} \beta \,, \\
e^a \wedge \left[ \alpha \cwedge{m} \left( X_a \hook \beta \right) \right] &=
  \left[ (-1)^\pi \alpha\right] \cwedge{m} \left[ (\pi-m)\beta \right] \, .
\end{aligned}
\end{equation}

When dealing with the Dirac equation we use the following isomorphism $\gamma_*$ between $\Omega(\M)$ and the Clifford bundle: 
\ba 
f &=& \frac{1}{p!} f_{\mu_1 \dots \mu_p} dx^{\mu_1} \dots dx^{\mu_p} \mapsto \nn \\ 
&& \gamma_* (f) = \frac{1}{p!} f_{\mu_1 \dots \mu_p} \Gamma^{\mu_1 \dots \mu_p} \, . 
\ea 
Here $\{ \Gamma^\mu \}$ are the Gamma matrices, satisfying the standard relation $\Gamma^\mu \Gamma^\nu + \Gamma^\nu \Gamma^\mu = 2 g^{\mu\nu}$,  $\Gamma^{\mu_1 \dots \mu_p} = \Gamma^{[\mu_1} \dots \Gamma^{\mu_p]} $  and the equation above straightforwardly generalises to the case of an inhomogeneous form. Any time the context makes it clear that quantities refer to the Clifford bundle, we will write $f$ instead of $\gamma_* (f)$. For example under these conditions the Dirac operator is written as $D= e^a \nabla_a$. The product of two Clifford bundle forms can be re-expressed in terms of contracted wedge products using the Gamma matrix algebra. Let $\alpha\in\Omega^p(M)$,  $\beta\in\Omega^q(M)$ and $p\leq q$. Then the Clifford product expands as
\be\label{usefulProduct1}
\alpha \beta = \sum_{m=0}^p \frac{ (-1)^{m(p-m) + [m/2]}}{m!} \alpha \cwedge{m}  \beta  \, , 
\ee
and
\be \label{usefulProduct2}
\beta \alpha = (-1)^{pq} \sum_{m=0}^p \frac{ (-1)^{m(p-m +1) + [m/2]}}{m!} \alpha \cwedge{m}  \beta  \, .
\ee

\subsection{Basics of hidden symmetries} 
Hidden symmetries of a Hamiltonian physical system are associated to conserved quantities of the dynamics that are polynomial in the momenta. If the system is classical by momenta we mean the variables $p_\mu$ canonically conjugated to the position variables $x^\mu$, and if the system is quantum mechanical we mean the appropriate operators. When the spacetime admits a Killing vector  $K$ then the conserved quantity is  of order one in the momenta and vice-versa, if there is such a quantity it can be written in terms of a Killing vector and its derivatives. Of particular importance are the following two classes of special tensors. 

{\it Killing-St\"{a}ckel} tensors (KS) are symmetric tensors $K^{\mu_1 \dots \mu_p} = K^{(\mu_1 \dots \mu_p)}$ satisfying the differential equation 
\be 
\nabla^{( \lambda} K^{\mu_1 \dots \mu_p)} = 0 \, .  
\ee 
They generate conserved quantities 
\be 
C_K = K^{\mu_1 \dots \mu_p} p_{\mu_1} \dots p_{\mu_p} 
\ee 
for the theory of the classical free scalar particle in curved space. A well known example is given by Carter's constant for the Kerr metric, and Carter like constants for curved backgrounds keep being discovered in recent research \cite{Santillan2012}. 
Quantum mechanically, the corresponding operators in the case of rank $2$ are given by 
\be 
\mathcal{K} =  \nabla_\mu \left[ K^{\mu\nu} \nabla_\nu \right] \, , 
\ee 
but these do not always generate conserved quantities, as the commutator $[\mathcal{K} , \nabla_\mu \nabla^\mu ]$ is given by an appropriate contraction of the Ricci tensor with $K$ \cite{Carter1977}. Failure of the commutator to close on zero indicates that the classical symmetry is gravitationally anomalous. 
If the spacetime is special, for example in the case of the Kerr-NUT-(A)dS spacetime, then the anomaly vanishes. Another special case when the anomaly vanishes is when the KS tensor can be written as the square of a Killing-Yano tensor, which will be defined below. 
In the case of rank $2$ KS tensors, the theory of the supersymmetric spinning particle admits a superfield that is the generalisation of the phase space function $K^{\mu\nu} p_\mu p_\nu$, a candidate conserved quantity that is also supersymmetric. In this case too in general there is an anomaly and the superfield is not supersymmetric nor conserved, but for Kerr-NUT-(A)dS spacetimes the anomaly vanishes \cite{DavidMarco2011}. Finally, it is worth noticing that a Killing vector is a Killing-St\"{a}ckel tensor of rank $1$. 
 
{\it Conformal Killing-Yano} tensors (CKY) are forms $\omega_{\mu_1 \dots \mu_p} = \omega_{[\mu_1 \dots \mu_p]}$ such that 
\be 
\nabla_\lambda \, \omega_{\mu_1 \dots \mu_p} = \nabla_{[\lambda} \,  \omega_{\mu_1 \dots \mu_p]} + \frac{p}{D-p+1} g_{\lambda [\mu_1} \nabla^\rho \omega_{|\rho| \mu_2 \dots \mu_p]} \, , 
\ee
or equivalently without using components 
\be \label{eq:CKY_definition}
\nabla_X \omega = \frac{1}{\pi +1} X \hook d \omega - \frac{1}{\dm - \pi +1} X^\flat \wedge \delta \omega \, , 
\ee
for any vector $X$. This formula generalises automatically to the case of inhomogeneous forms. 
When $\omega$ is co-closed, $\delta \omega = 0$, $\omega$ is a Killing-Yano form (KY), and when it is closed, $d\omega =0$, it is a closed conformal Killing-Yano form (CCKY). Equation \eqref{eq:CKY_definition} is invariant under Hodge duality, interchanging KY and CCKY tensors. 
Benn, Charlton, and Kress \cite{BennCharlton:1997, BennKress:2004} have shown the important result that, in all dimensions  $\dm$ and arbitrary signature, first-order symmetry operators of the massless Dirac operator are in one to one correspondence with CKY forms. Specifically, if $S$ is  an operator satisfying $DS=RD$ for some operator ${R}$, then $S$ is given by 
is given by
\be
S=S_\omega+\alpha D\,,
\ee
where $\alpha$ is an arbitrary inhomogeneous form, and $S_\omega$, given in terms of an inhomogeneous CKY form $\omega$ obeying \eqref{eq:CKY_definition},  is
\be\label{SOProp1}
S_\omega=X^a\hook\omega\, \nabla_a+\frac{\pi-1}{2\pi}d\omega-\frac{n-\pi-1}{2(n-\pi)}\delta\omega\;.
\ee
Then $S_\omega$ obeys
\be \label{eq:CKY_graded_commutator}
\bgc{D}{S_\omega}\equiv DS_\omega-(\eta S_\omega)D=-\Bigl(\frac{\eta}{n-\pi} \delta\omega\Bigr)D\,.
\ee
The freedom of adding an arbitrary form $\alpha$ is unavoidable. It can also be shown that if $\omega$ is a CCKY form then the operator 
\be \label{eq:symmetry_operator_CCKY}
S_\omega=e^a \wedge \omega\, \nabla_a -\frac{n-\pi-1}{2(n-\pi)}\delta\omega 
\ee 
either commutes or anti-commutes with the Dirac operator, depending whether $\omega$ is even or odd \cite{MDP2011_1}.

Similar results hold in the case of the spinning particle, see  for example \cite{GaryetAl1993} for the discussion of KY tensors. For the Kerr-NUT-(A)dS metrics it is possible to show that there exist $\dm$ independent such operators, as many as the number of dimensions (one of them being $D$ itself),  that they all mutually commute and that this explains the separation of variables for the Dirac equation in these metrics \cite{MDP2011_1,MDP2011_2}. 

If a spacetime admits Killing spinors, with or without torsion, then these can be used to build CKY tensors \cite{Semmelmann2002,Marco2004,DavidetAl2012}.  The Dirac equation with skew-symmetric torsion has been discussed in \cite{David_withTorsion}. The link between CKY tensors and G--structures has been discussed in \cite{Papadopoulos1,Papadopoulos2,Santillan}.

\section{The Eisenhart-Duval lift\label{sec:Eisenhart}} 

\subsection{The geometric lift} 
In this paper we will consider the Eisenhart-Duval lift in the time independent case. Let $\mathcal{M}$ be a $\dm$ dimensional spacetime, with metric 
\be 
g = g_{\mu\nu} (x) dx^\mu dx^\nu \, . \label{eq:lower_dim_h}
\ee 
with Euclidean signature. 
On $\M$ we can consider the classical theory of a particle of mass $m$ and electric charge $e$, interacting with a position dependent potential $V(x)$ and with a stationary electromagnetic field with vector potential $A_\mu(x)$, introducing the Hamiltonian 
\be 
\mathcal{H} = \frac{1}{2m} g^{\mu\nu} \left(p_\mu - e A_\mu \right) \left(p_\nu - e A_\nu \right) + V(x) \, , 
\ee 
where $p_\mu$ is the canonical momentum. The Hamiltonian function written above is not explicitly invariant  invariant under a gauge transformation of the vector potential but the full theory is. What happens is that under a gauge transformation the canonical momenta change as well as the vector potential, generating a canonical transformation. As a result the Hamiltonian equations of motion are not explicitly gauge invariant but the theory in fact is. It is possible to introduce gauge invariant momenta $P_\mu = p_\mu - e A_\mu$ while at the same time modifying the Poisson brackets and recognising that $P$ acts as a $U(1)$ covariant derivative, see \cite{GaryetAl1993,JackiwManton1980,DuvalHorvathy2004,vanHolten2006,HorvathyNgome2009,Visinescu2011}. This is not needed to the extent of the calculations done in this paper, and we will work with the canonical $p_\mu$ variables. 

It is a result by Eisenhart \cite{Eisenhart1928} that the Hamiltonian \eqref{eq:lower_dim_h} can be obtained by reduction from the following Hamiltonian in $\dm +2$ dimensions: 
\ba
\hat{\mathcal{H}} = \og^{MN} \h{p}_M \h{p}_N &=& g^{\mu\nu} \left(p_\mu - \frac{e}{m} A_\mu p_v \right) \left(p_\nu - \frac{e}{m} A_\nu p_v \right) \nn \\ 
&& + 2 p_v p_t + \frac{2}{m}V p_v^2 \, , 
\ea 
where $\h{p}_M = (p_v, p_t, p_{\mu_1}, \dots, p_{\mu_\dm})$. $\hat{\mathcal{H}}$ describes the motion of a massless particle in the higher-dimensional Lorentzian metric 
\ba  \label{eq:Eisenhart_metric}
\h{g} &=& \og_{MN} d \h{x}^M d \h{x}^N \nn \\ 
&&  \hspace{-0.75cm} =  g_{\mu\nu} dx^\mu dx^{\nu} + \frac{2e}{m} A_\mu dx^\mu dt + 2 dt dv - \frac{2}{m}V dt^2 \, . 
\ea 
To see this consider the coordinate $v$, generated by the covariantly constant Killing vector $\xi = \partial/\partial v$, which is conjugate to $p_v$. $p_v$ is constant along a solution of the equations of motion. If we specialise to a null solution with $\hat{\mathcal{H}} = 0$ and choose $p_v =m$ then we have 
\be 
0 = \hat{\mathcal{H}} = g^{\mu\nu} \left(p_\mu - e A_\mu  \right) \left(p_\nu - e A_\nu \right) + 2 m p_t + 2m V \, , 
\ee 
or equivalently 
\be 
 p_t = - \mathcal{H} \, . 
\ee 
This means we can identify $\mathcal{H}$, generator of time translations in the $\dm$-dimensional system, with $-p_t$ in the $\dm+2$-dimensional one, which generates translations along $-\partial/\partial t$.  Expressions for the vielbeins of the metric \eqref{eq:Eisenhart_metric}, as well as the dual vector base, the Levi-Civita connection and spin connection components  can be found in appendix \ref{apdx:Eisenhart_metric}. 

Geometrically we can describe $\h{\M}$ as a bundle over $\M$, with projection $\Pi: (t,v, x^\mu) \mapsto x^\mu$. Then if $f$ is a $p$--form defined on $\M$ its pull-back on $\h{\M}$ under the map $\Pi^*$ is a $p$--form $\h{f}$ on $\h{\M}$. 
 
As seen above null geodesics on $\h{\M}$ relative to $\h{g}$ generate massive geodesics on $\M$ relative to $g$. It is in fact possible to do more: given a generic conserved quantity for the motion on $\M$ that is a non-homogeneous polynomial in momenta, in other words a hidden symmetry, this can be lifted to an appropriate hidden symmetry on $\h{\M}$ that is homogeneous in momenta and that therefore is associated to a Killing tensor. The Poisson algebra on $\M$ of conserved charges for the original motion then is the same as the Schouten-Nijenhuis algebra of the Killing tensors associated to lifted conserved charges  \cite{DavidGaryetAl2011}. This means that the dynamical evolution on $\M$ as described in full phase space can be embedded in the higher dimensional phase space. As we will see this does not happen in the case of the Dirac equation, where in general it will not be possible to lift all hidden symmetries from $\M$ to $\h{\M}$. This corresponds to the fact that when performing the dimensional reduction of the Dirac equation on $\h{\M}$ a non-trivial projection is required in phase space in order to recover the Dirac equation with $V$ and $A$ flux on $\M$. This projection is not compatible with all hidden symmetry transformations.

\subsection{Lift of conformal Killing-Yano forms\label{sec:lift}}
In this section we consider the CKY equation on $\h{\M}$: 
\be \label{eq:CKY}
\onabla_{\oX} \of = \frac{1}{\pi +1} \oX \hook \hat{d} \of - \frac{1}{(\dm +2) - \pi +1} \oX^\flat \wedge \hat{\delta} \of \, , 
\ee 
$\forall \, \h{X}$ vector. We will specialise to a homogeneous form $\h{f}$, since any non-homogeneous CKY form can be split into a sum of homogeneous CKY forms. CKY forms are the appropriate forms to look for, since in $\dm +2$ dimensions we are focussing on null geodesics that are in correspondence to geodesics on the base manifold $\M$, and since they generate symmetries of the massless Dirac equation. 
 
Before studying the general case  we begin with four simplified anst\"{a}tze for the higher dimensional CKY form. Given a  $p$--form $f = f(x)$ living on the base manifold $\M$ we can build the following higher dimensional forms: 
\ba 
\h{f}_1 &=& f \, ,  \label{eq:ansatz1} \\ 
\h{f}_2 &=& e^+ \wedge f \, , \label{eq:ansatz2} \\ 
\h{f}_3 &=& e^- \wedge f \, , \label{eq:ansatz3} \\ 
\h{f}_4 &=& e^+ \wedge e^- \wedge f \, , \label{eq:ansatz4} 
\ea 
where by writing on the right hand side $f$ instead of $\h{f}$ for a form on $\h{M}$ we are performing a slight abuse of notation with the purpose of indicating that $f$ represents the canonical embedding in $\h{\M}$ of a form originally defined on $\M$. 
Since in principle each of the four forms above can be multiplied times a function of the $t$ and $v$ variables, we allow from the beginning for a $p$--form $\h{f}_i = \h{f}_i \, (v,t,x)$ that can have $v, t$ dependence. Hodge duality in $\oM$ maps a form of type $\h{f}_1$ into one of type $\h{f}_4$ - after allowing for $f \rightarrow *_{\M} f$. Similarly, it relates the forms $\h{f}_2$ and $\h{f}_3$ to themselves. 
 
We will first study the conditions under which these four ans\"{a}tze generate CKY tensors on $\h{\M}$. After that we will study the equations for the general CKY tensor. Two main findings are worth noticing. One is that in case of eqs.\eqref{eq:ansatz1} and \eqref{eq:ansatz4} it is possible to generate KY and, respectively, CCKY tensors on $\h{\M}$ when $f$ is KY and, respectively CCKY on $\M$. By doing this we can construct new examples of Lorentzian metrics with CKY tensors by lifting known CKY tensors in Riemannian signature, for example when $\M$ is the Kerr-NUT-(A)dS metric or the Taub-NUT metric \cite{GaryRuback1987,FeherHorvathy1988,VamanVisinescu1997,BaleanuCodoban1998,Visinescu2000}. Secondly, it will not be possible to lift a generic CKY tensor on $\M$ to a CKY tensor on $\h{\M}$. This will be discussed more in detail in section \ref{sec:Dirac}, where it will be shown that the process of lift/oxidation and its inverse, reduction, at the level of the Dirac equation involves a non-trivial projection in phase space, and this is not compatible with all lower and higher dimensional hidden symmetries.

\subsubsection{Ansatz 1\label{sec:ansatz1}}
Consider the $p$--form $\hat{f}_1$ on $\oM$ given by \eqref{eq:ansatz1}. Let's check under which conditions this satisfies the $\dm +2$ dimensional CKY equation \eqref{eq:CKY}. 

The CKY equation \eqref{eq:CKY} splits into three types of equation, one for each of $\oX = \oX^+$, $\oX = \oX^-$, and $\oX = \oX^a$. We can analyse each using eqs. \eqref{eq:nabla_minus_f}, \eqref{eq:nabla_plus_f},  \eqref{eq:nabla_a_f} and \eqref{eq:df}, \eqref{eq:delta_f}. 
The $\oX = \oX^+$ component gives 
\be 
\partial_v f = 0 \, , 
\ee 
\be \label{eq:ansatz1_useful0}
\delta f = 0 \, ,  
\ee
the $\oX = \oX^-$ component 
\be \label{eq:ansatz1_useful_dV}
 dV^\sharp \hook f  = 0 \, ,  
\ee
\be \label{eq:ansatz1_useful1}
F \cwedge{2} f = 0  \, , 
\ee 
\be \label{eq:simple_ansatz1_t_derivative}
\partial_t  f +  \frac{p+1}{p}  \frac{e}{2m} F \cwedge{1} f = 0 \, , 
\ee 
and lastly, the $\oX = \oX^a$ component gives  
\be 
\frac{e}{2m} (X_a \hook F) \cwedge{1} f =  \frac{1}{p+1} X_a \hook \partial_t f \, ,
\ee 
\be \label{eq:ansatz1_KY}
\nabla_a f = \frac{1}{p+1} X_a \hook df   \, .  
\ee 
The latter equation is the Killing-Yano equation on the base. The former instead implies \eqref{eq:simple_ansatz1_t_derivative}. 

Thus we have a $t$-parameterised family of KY forms on $\M$. But this is compatible with eq.\eqref{eq:simple_ansatz1_t_derivative} only if $F \cwedge{1} f$ is KY as well, which in general will not be the case. Then it must be that separately 
\be 
\partial_t f = 0 \, , 
\ee 
\be 
(X_a \hook F) \cwedge{1} f \, , 
\ee 
which implies
\be \label{eq:ansatz1_useful2}
F \cwedge{1} f = 0 \, . 
\ee 
Thus there is no $v$, $t$ dependency and we discover that a $KY$ form on the base manifold $\M$ can be lifted directly to a KY form on $\oM$, since the conditions found imply $\h{\delta} \h{f}_1 = 0$. With such a form we can construct a symmetry operator for the Dirac equation on $\h{\M}$, and when $p$ is odd we know that such operator strictly commutes with the Dirac operator $\h{D}$ \cite{MDP2011_1}. Also for such values of $p$ the conditions we have found in this section guarantee that on $\M$ we can build a symmetry operator for the Dirac equation with $V$ and $A$ flux \cite{DavidClaudePavel2011}. In section \ref{sec:Dirac} we show how in the case of $p$ odd it is possible to dimensionally reduce such hidden symmetry operator on $\h{\M}$ to get a hidden symmetry operator associated to flux on $M$. We will also see how this is not possible if $p$ is even, which goes in agreement with the fact that the conditions required on an even CKY tensor with flux in \cite{DavidClaudePavel2011} are different. Such other conditions are those to be found in sec.\ref{sec:ansatz4}. It is worthwhile realising that the conditions found here are more restrictive than those in \cite{DavidClaudePavel2011}, so  in general it is possible to conceive the existence of tensors that satisfy the less restrictive conditions, thus generating symmetries of the Dirac equation with flux on $\M$, but that at the same time do not satisfy the conditions of this section and therefore cannot be lifted to $\h{\M}$. 

Lastly, we notice that since Hodge duality exchanges KY with CCKY forms, the result of this section implies that the form $\h{f}_4$ of equation \eqref{eq:ansatz4} is expected to be a CCKY form on $\h{\M}$, with the $f$ function appearing there a CCKY form on $\M$.

\subsubsection{Ansatz 2\label{sec:ansatz2}}
Consider the $p+1$--form $\hat{f}_2$ on $\oM$ given by 
\be 
\hat{f_2} = \h{e}^+ \wedge f \, . 
\ee 
We can calculate derivatives of $\h{f}_2$ and get 
\be 
\h{\nabla}_- \h{f}_2 = \h{e}^+ \wedge \partial_v f \, , 
\ee
\be 
\h{\nabla}_+ \h{f}_2 = \h{e}^+ \wedge \left[ \left(  \frac{V}{m} \partial_v + \partial_t \right) f +  \frac{e}{2m} \left(F \cwedge{1} f \right) \right] \, , 
\ee  
\be 
\h{\nabla}_a \h{f}_2 = \h{e}^+ \wedge \left( - \frac{e}{m} A_a \partial_v f + \nabla_a f \right) \, , 
\ee 
\be 
\h{d}\h{f}_2  = - \h{e}^+ \wedge \h{e}^- \wedge \partial_v f + \h{e}^+ \wedge \left( \frac{e}{m} A \wedge \partial_v f - df\right) \, , 
\ee 
\be 
\h{\delta}\h{f}_2  = - \partial_v f + \h{e}^+ \wedge \left( - \frac{e}{m} A^\sharp \hook \partial_v f - \delta f \right) \, . 
\ee 
The $\oX = \oX^+$ component of the CKY equation gives 
\be 
\partial_v \hat{f} = 0 \, , 
\ee 
the $\oX = \oX^-$ component 
\be 
\delta \h{f} = 0 \, , 
\ee 
\be 
d \h{f} = 0 \, , 
\ee 
\be 
\partial_t \h{f} + \frac{e}{2m} F \cwedge{1} \h{f} = 0 \, ,  
\ee 
and lastly the $\oX = \oX^a$ component gives  
\be 
\nabla_a \h{f} = 0 \, . 
\ee
So in this case $f$ and $\h{f}_2$ are covariantly constant forms.

\subsubsection{Ansatz 3\label{sec:ansatz3}}
Consider the $p+1$--form $\hat{f}$ on $\oM$ given by 
\be 
\hat{f}_3 = \h{e}^- \wedge f \, . 
\ee 
The explicit form of the derivatives of $\h{f}_3$ is: 
\be 
\h{\nabla}_- \h{f}_3 = \h{e}^- \wedge \partial_v f \, , 
\ee
\ba 
\h{\nabla}_+ \h{f}_3 &=& \h{e}^- \wedge \left[ \left(  \frac{V}{m} \partial_v + \partial_t \right) f  + \frac{e}{2m} F \cwedge{1} f \right]  \nn \\ 
&& + \h{e}^+ \wedge \h{e}^- \wedge \left( \frac{dV^\sharp}{m} \hook f \right) + \frac{dV}{m} \wedge f \, , 
\ea 
\ba 
\h{\nabla}_a \h{f}_3 &=& \h{e}^- \wedge \left( - \frac{e}{m} A_a \partial_v f + \nabla_a f \right) \nn \\ 
&& + \frac{e}{2m} \h{e}^+ \wedge \h{e}^- \wedge \left[ \left(X_a \hook F \right) \cwedge{1} f \right] \nn \\ 
&& + \frac{e}{2m} \left(X_a \hook F \right) \wedge f \, , 
\ea 
\ba
&& \h{d}\h{f}_3  =  \h{e}^+ \wedge \frac{dV}{m}  \wedge f + \h{e}^- \wedge \left( \frac{e}{m} A \wedge \partial_v f - df \right)  \nn \\ 
&& + \h{e}^+ \wedge \h{e}^- \wedge \left( \frac{V}{m} \partial_v + \partial_t \right) \hat{f}  + \frac{e}{m} F \wedge f \, , 
\ea 
\ba 
\h{\delta}\h{f}_3  &=& \h{e}^+ \wedge \left( \frac{dV^\sharp}{m}  \hook f \right) + \h{e}^- \wedge \left( - \frac{e}{m} A^\sharp \hook \partial_v f - \delta f \right)  \nn \\ 
&& + \h{e}^+ \wedge \h{e}^- \wedge \left( \frac{e}{2m} F \cwedge{2} f \right) \nn \\ 
&& -  \left[ \left( \frac{V}{m} \partial_v + \partial_t \right) \hat{f} + \frac{e}{2m} F \cwedge{1} f \right]  \, . 
\ea 
The CKY equation gives again a covariantly constant case, with $\partial_v f = 0 = \partial_t f$, $\nabla_a f = 0$.

\subsubsection{Ansatz 4\label{sec:ansatz4}}
Lastly consider the $p+2$--form $\hat{f}_4$ on $\oM$ given by 
\be 
\hat{f}_4 = \h{e}^+ \wedge \h{e}^- \wedge f \, .  
\ee
Its derivatives are given by 
\be 
\h{\nabla}_- \h{f}_4 = \h{e}^+ \wedge \h{e}^- \wedge \partial_v f \, , 
\ee
\ba 
&& \h{\nabla}_+ \h{f}_4 = \h{e}^+ \wedge \frac{1}{m} dV \wedge f \nn \\ 
&& + \h{e}^+ \wedge \h{e}^- \wedge \left[ \left( \frac{V}{m} \partial_v + \partial_t \right) f + \frac{e}{2m} F \cwedge{1} f \right] \, , 
\ea  
\ba  
\h{\nabla}_a \h{f}_4 &=& \h{e}^+ \wedge \frac{e}{2m} \left( X_a \hook F \right) \wedge f \nn \\ 
&& + \h{e}^+ \wedge \h{e}^- \wedge \left[ - \frac{e}{m} A_a \partial_v f + \nabla_a f \right]  \, , 
\ea  
\be 
\h{d}\h{f}_4  = - \h{e}^+ \wedge \frac{e}{m} F \wedge f + \h{e}^+ \wedge \h{e}^- \wedge \left( - \frac{e}{m} A \wedge \partial_v f+ df\right) \, , 
\ee 
\ba 
\h{\delta}\h{f}_4  &=& - \h{e}^- \wedge \partial_v f + \h{e}^+ \wedge \left( \frac{V}{m} \partial_v + \partial_t \right) f \nn \\ 
&& + \h{e}^+ \wedge \h{e}^- \wedge \left( \frac{e}{m} A^\sharp \hook \partial_v f + \delta f \right)   \, . 
\ea
These are exactly the Hodge dual of the equations for $\h{d}\h{f}_1$ and $\h{\delta}f_1$ found in section \ref{sec:ansatz1}, with the understanding that the form $f$ there is related to the $f$ form of this section by Hodge duality on the base manifold $\M$. The CKY equations therefore lead to the Hodge dual of the conditions found there, as it can be checked using the identities in appendix \ref{apdx:Hodge_duality}, namely: 
\be 
\partial_{v,t} f = 0 \, , 
\ee 
\be 
d f = 0 \, , 
\ee 
\be 
dV \wedge f = 0 \, , 
\ee 
\be 
(X_a \hook F ) \wedge g = 0 \, , 
\ee 
\be 
F \wedge f = 0 \, 
\ee 
\be 
F \cwedge{1} f = 0 \, 
\ee  
\be \label{eq:ansatz4_CKY}
\nabla_a f = - \frac{1}{\dm - p +1} e_a \wedge \delta f \, . 
\ee 
There are two things worth noticing. The first is that eq.\eqref{eq:ansatz4_CKY} is the CCKY equation on the base manifold, and that this is ultimately made possible by the fact that $\dm - p + 1 = (\dm +2) - (p+2) + 1$. When all the conditions are satisfied $\h{f}_4$ is a CCKY tensor on $\h{\M}$. The second thing worth noticing is that the remaining conditions  guarantee that if $p$ is even we can build a symmetry operator of the Dirac equation on $\oM$ and of the Dirac equation with $V$ and $A$ flux on $\M$ \cite{DavidClaudePavel2011}. These conditions complement those found in section \ref{sec:ansatz1}, and are associated to the dimensional reduction to $\M$ of a symmetry operator on $\h{\M}$, as will be discussed in more detail in sec.\ref{sec:Dirac}. In this case too the conditions are stronger than those found in \cite{DavidClaudePavel2011}.

\subsubsection{The general CKY tensor} 
We are now ready to tackle the general case. Let the $p$--form $\h{f}$ be parameterised as 
\be \label{eq:CKY_most_general_param}
\hat{f} = f + \eh^+ \wedge \rho^+ + \eh^- \wedge \rho^- + \eh^+ \wedge \eh^- \wedge g \, , 
\ee
where the forms $f, \rho^{\pm}, g$ are, respectively, $p$, $p-1$ and $p-2$ forms defined on $\M$.  We can calculate all its derivatives by adding the derivatives calculated in the four previous ans\"{a}tze. The full conditions obtained from the CKY equation are somewhat long and are listed in appendix \ref{apdx:GeneralCase}. We are interested here in discussing the special case where the forms $f$, $\rho^\pm$, $g$ do not depend explicitly on the $v$, $t$ coordinates, or in other words the Lie derivative of $\h{f}$ with respect to the Killing vectors $\partial_v$ and $\partial_t$ is zero. In this case we can relate the parameterisation \eqref{eq:CKY_most_general_param} to the lift of a set of forms defined on the base manifold $\M$, while in the most general case listed in the appendix it is necessary to consider $v$ and $t$ parameterised families of forms. 

In general the CKY equation reduce to a series of equations on $\M$. We label these according to two indices $i$ and $J$ according to the following convention. The first index $i = -,+,a$ reflects which derivative is used in the CKY equation among $\h{\nabla}_-$, $\h{\nabla}_+$, $\h{\nabla}_a$. Once the appropriate CKY equation is chosen, the second index $J$, taking values $J= +, - , \pm, a$, is related to taking the projection on $\M$ of the CKY equation along the $\h{e}^+$, $\h{e}^-$, $\h{e}^+ \wedge \h{e}^-$ and $\h{e}^a$ direction. The full set of equations is 
\ba
&& (-+)  \; dg = 0  ,  \nn \\ 
&& (-+)^\prime \; \delta f = 0; \nn \\ 
&& (+-)  \;  F \cwedge{1} \rho^-  =0 \, , \nn \\
&& (++) \;  -\frac{dV^\sharp}{m} \hook f + \frac{dV}{m} \wedge g + \frac{e}{2m} F \cwedge{1} \rho^+ = 0  \, , \nn \\
&& (+ \pm) \; \frac{e}{2m} F \cwedge{1} g =  + \frac{1}{\dm + 3 - p} \left[ \frac{e}{2m} F \cwedge{2} f  - \delta \rho^+ \right. \nn \\ 
&& \left. - (\dm +2 - p) \frac{dV^\sharp}{m} \hook \rho^-   \right] = 0 \, , \nn \\
&& (+a) \;   \frac{e}{2m} F \cwedge{1} f = +  \frac{1}{p+1} \left[  - d \rho^+  - p \frac{dV}{m} \wedge \rho^-  - \frac{e}{m} F \wedge g \right] \, ,\nn \\ 
&& (a-) \;  \nabla_a \rho^- = 0 \, , \nn \\ 
&& (a+) \; - \frac{e}{2m} (X_a \hook F) \cwedge{1} f  + \nabla_a \rho^+ + \frac{e}{2m} (X_a \hook F) \wedge g = \nn \\ 
&&   \frac{X_a}{p+1} \hook \left[   d \rho^+  - \frac{dV}{m} \wedge \rho^-  + \frac{e}{m} F \wedge g \right] \nn \\ 
&&  + \frac{e_a}{\dm + 3 - p} \wedge \left[ \frac{e}{2m} F \cwedge{2} f  - \delta \rho^+ + \frac{dV^\sharp}{m} \hook \rho^-   \right] \, , \nn \\
&& (a \pm) \; \nabla_a g + \frac{e}{2m} (X_a \hook F) \cwedge{1} \rho^-   = \nn \\ 
&&    - \frac{e_a}{\dm + 3 - p} \wedge \left[ \delta g + \frac{e}{2m} F \cwedge{2} \rho^-    \right] \, , \nn \\
&& (ab) \;  \nabla_a f + \frac{e}{2m} (X_a \hook F) \wedge \rho^- = \nn \\ 
&&  \frac{X_a}{p+1} \hook \left[   + df + \frac{e}{m} F \wedge \rho^- \right]    \, . 
\ea  
These equations classify the most general CKY tensor on $\h{\M}$ that has zero Lie derivative with respect to $\partial_v$ and $\partial_t$ in terms of forms on $\M$. The last two equations are respectively a CCKY and a KY equation for $g$ and, respectively, $f$, with a deformation parameterised by the covariantly constant form $\rho^-$ and by $F$. The form $\rho^+$ instead is related to a deformation of the  equations with no derivatives for $f$ and $g$ that we have found in the previous sections for terms of the kind $dV \wedge$, $dV \hook$, $F \cwedge{1}$. 

In particular it can be seen that under no circumstance the form $f$ can be strictly CCKY with $\delta f \neq 0$ or the form $g$ can be strictly KY with $dg \neq 0$, thus proving that if any such form exists on $\M$ then it cannot be lifted to $\h{\M}$ in order to give a CKY form. The forms $\rho^\pm$, $f$ and $g$ satisfy a set of generalised interdependent CKY equations.

\subsubsection{Conditions for rank 2 CCKY tensors} 
It is of particular interest to discuss rank 2 CCKY tensors in the Eisenhart-Duval lift metrics given the fact there exists a classification of such tensors for Riemannian spacetimes. Therefore an interesting question is whether Eisenhart-Duval spacetimes can provide new non-trivial examples of such forms. 
 
If the form $\h{f}$ of the previous section is of rank $2$, then $g$ is a function and $\rho^\pm$ are $1$--forms. Equation $(-+): dg=0$ then tells us that $g$ is constant. Asking that $\h{f}$ is closed amounts to the two conditions 
\ba 
&& df + \frac{e}{m} F \wedge \rho^- = 0 \, , \\ 
&& d\rho^+ - \frac{dV}{m}\wedge \rho^- + \frac{e}{m} g F = 0 \, , 
\ea 
which have as solutions 
\ba 
&& f = - \frac{e}{m} A \wedge \rho^- + d \Lambda^{(1)} \, , \\ 
&& \rho^+ = \frac{V}{m} \rho^- - \frac{eg}{m} A + d\Lambda^{(0)} \, , \label{eq:rank2_case_rho_plus}
\ea 
where $\Lambda^{(0)}$ and $\Lambda^{(1)}$ are a $0$-- and, respectively, a $1$--form to be found by solving the other equations. 

The full set of equations are complicated to solve. One might hope to obtain a simplification in the case $F=0$, $V\neq 0$. If $F=0$ then eq.(++) gives 
\be 
\frac{dV^\sharp}{m} \hook f = g \frac{dV}{m} \, . 
\ee 
Taking the hook contraction of this equation with $dV^\sharp$ we get $0 = g \frac{|\vec{\nabla}V|^2}{m}$, and since $\M$ has euclidean signature this means that either $g=0$ or $V= constant$. We take $g=0$. The rest of the equations simplify to 

\be
\begin{split}
& (-+)^\prime  \delta f = 0  ,  \\ 
& (++) \frac{dV^\sharp}{m} \hook f = 0 \, , \\
& (+ \pm) \quad  \delta \rho^+  = - \dm  \frac{dV^\sharp}{m} \hook \rho^-     \, , \\
& (+a) \quad    d \rho^+  = - 2 \frac{dV}{m} \wedge \rho^-    \, ,\\ 
& (a-) \quad  \nabla_a \rho^- = 0 \, , \\ 
& (a+)  \nabla_a \rho^+  =  \frac{e_a}{\dm +1} \wedge \left( - \delta \rho^+ + \frac{dV^\sharp}{m} \hook \rho^-  \right)\, , \\
& (ab) \quad  \nabla_a f = 0    \, . 
\end{split} 
\ee 
Now we can see that eq.$(a+)$ implies $d\rho^+ = 0$, and this together with $(+a)$ implies $dV \wedge \rho^- = 0$, or 
\be 
\frac{V}{m} = \alpha \rho^- \cdot \vec{x} + \beta \, , 
\ee 
where $\alpha$ and $\beta$ are constants. 

Now $(+\pm)$ together with $(a+)$ give 
\be
\nabla_a \rho^+  =  \left( \frac{dV^\sharp}{m} \hook \rho^- \right)  e_a = \alpha |\rho^-|^2 e_a  \, , 
\ee 
and this together with eq.\eqref{eq:rank2_case_rho_plus} yields the following equation for the form $\Lambda^{(0)}$: 
\be \label{eq:Lambda_0}
\nabla_a (d\Lambda^{(0)}) = - \alpha ( \rho^-_a \rho^-_b - |\rho^-|^2 \delta_{ab}) e^b \, . 
\ee
One might try to solve this equation when $\M$ is the Euclidean flat space. A solution can indeed be displayed, however in this case $\h{M}$ is flat, and therefore this corresponds to no new metrics. Regardless of the specific form of $\M$ the conditions we found so far guarantee that the co-differential of $\h{f}$ is a null form: 
\be 
\h{\delta}\h{f} = (\dm +1) \alpha (\rho^-)^2 e^+ \, .  
\ee
 
We do not pursue here the task of solving \eqref{eq:Lambda_0} when $\M$ is a non-flat space, or that of solving the full set of equations when $F\neq 0$. However we consider this an interesting task given its potential to generate new metrics with rank 2 CCKY tensors that are not covered by the Riemannian classification.

\section{Dirac equation\label{sec:Dirac}}
In this section we describe in detail the massless Dirac equation on $\h{\M}$ and show that in general it gives rise to a non-relativistic L\'evy-Leblond equation. We then specialise to those special cases when it is possible to reduce the higher dimensional Dirac equation and obtain again a Dirac equation in lower dimension.  First, we show how an appropriate non-trivial projection on the higher dimensional spinor allows to reduce the massless Dirac equation on $\h{\M}$ to the Dirac equation with $V$ and $A$ flux on $\M$. We also show how to perform the inverse lift operation, that is how to embed a lower dimensional solution of the Dirac equation with flux into a solution of the massless higher dimensional Dirac equation. Second, we consider the dimensional reduction of hidden symmetry operators associated to the KY and CCKY tensors of sections \ref{sec:ansatz1} and \ref{sec:ansatz4}, showing how only a subset of these commutes with projection operation and thus yields hidden symmetry operators for the theory on $\M$.

\subsection{Dimensional reduction and lift} 
 
\subsubsection{Non-relativistic L\'evy-Leblond equation} 
In \cite{HorvathyZhang2009} it has been shown how the massless free Klein-Gordon equation in $d+2$--dimensional Lorentzian spacetime can be reduced to the massive Schr\"{o}dinger equation in Riemannian $d$--dimensional spacetime using a projection on the base space $\M$ of the Eisenhart-Duval spacetime $\h{\M}$. In this section  we show analogously how dimensional reduction of the massless Dirac equation on $\h{\M}$ yields its non-relativistic counterpart on $\M$, that is the L\'evy-Leblond equation \cite{LevyLeblond1967}. The first derivation of the L\'evy-Leblond equation from a lightlike reduction from 4 and 5 dimensions has been given in \cite{DuvalHorvathyPalla1996}. 

Spinors on $\M$ have dimension $2^{\left[ \frac{\dm}{2} \right]}$, while spinors on $\oM$ have dimension $2^{\left[ \frac{\dm +2}{2} \right]}= 2 \cdot 2^{\left[ \frac{\dm}{2} \right]}$. Given the Pauli matrices 
\begin{equation}\label{eq:Pauli_matrices}
\begin{gathered}
\sigma_1 \equiv
    \left(\begin{array}{cc}
        0 & 1 \\
        1 & 0 \\
      \end{array}\right)
    \;,\quad
    \sigma_2 \equiv
    \left(\begin{array}{cc}
        0 & -i \\
        i & 0 \\
      \end{array}\right)
    \; , \quad     
\sigma_3 \equiv
    \left(\begin{array}{cc}
        1 & 0 \\
        0 & -1 \\
      \end{array}\right)
    \; ,  
\end{gathered}
\end{equation}
we define $\sigma^\pm = \frac{\sigma_1 \pm i \sigma_2}{2}$. These satisfy $(\sigma^\pm)^2 = 0$, $\left\{ \sigma^+ , \sigma^- \right\} = \mathbb{I}$. Let $\Gamma^a$ be gamma matrices for $\M$, and $\h{\Gamma}^A$ for $\oM$. We use the following representation for gamma matrices on $\h{\M}$: 
\be \label{eq:gamma_matrices_representation} 
\left\{ \begin{array}{rcl} 
\h{\Gamma}^+ &=& \sigma^+ \otimes \mathbb{I} \, ,  \\ 
\h{\Gamma}^- &=& \sigma^- \otimes \mathbb{I} \, , \\ 
\h{\Gamma}^a &=& \sigma_3 \otimes \Gamma^a \, . 
\end{array} \right. 
\ee 
Using the explicit form of the spin connection \eqref{eq:spin_connection} we find the expression for the covariant derivatives of a spinor $\h{\psi}$ on $\oM$: 
\be 
\left\{ \begin{array}{rcl} 
\h{\nabla}_- &=& \partial_v \, ,  \\ 
\h{\nabla}_+ &=& \left(\frac{V}{m} \partial_v + \partial_t \right) - \frac{1}{2m}\h{\Gamma}^+ dV - \frac{e}{4m} F \, , \\ 
\h{\nabla}_a &=& \nabla_a - \frac{e}{m} A_a \partial_v + \frac{e}{4m} \h{\Gamma}^+ \h{\Gamma}^b F_{ba}  \, . 
\end{array} \right. 
\ee
Thus we can write the Dirac operator on $\oM$ as 
\ba 
\h{D} = \h{\Gamma}^A \h{\nabla}_A &=&  \h{\Gamma}^- \partial_v + \h{\Gamma}^+ \left( \frac{V}{m}\partial_v + \partial_t + \frac{e}{4m} F \right) \nn \\ 
&& + \h{\Gamma}^a \left( \nabla_a - \frac{e}{m} A_a \partial_v \right) \, . 
\ea 
In the case of a Killing vector $\h{K}$ the symmetry operators of eq.\eqref{SOProp1} assume the form 
\be 
\h{S}_{\h{K}} = \h{\nabla}_{\h{K}} + \frac{1}{4} \h{d} \h{K} \, . 
\ee 
For the two Killing vectors $\oX^+$ and $\oX^- - \frac{V}{m} \oX^+$ of the lift metric the operators can be calculated  explicitly and are given by 
\be 
K_{\oX^+} = \partial_v \, , 
\ee 
and 
\be 
K_{(\oX^- - \frac{V}{m} \oX^+)} = \partial_t \, . 
\ee
They both commute with the Dirac operator $\h{D}$ on $\oM$, so we can ask that a spinor $\h{\psi}$ on $\oM$ that satisfies $\h{D} \h{\psi} = 0$ is also an eigenspinor of the two operators. For the purpose of recovering the L\'evy-Leblond equation we ask the less restrictive condition of $\h{\psi}$ being eigenspinor only of $K_{\h{X}^+}$: 
\be 
\partial_v \h{\psi} = im \h{\psi} \, . 
\ee
We have chosen the $\partial_v$ eigenvalue to be proportional to the mass parameter $m$ in order for the Dirac operator to reduce to 
\be 
\h{D} = i m  \h{\Gamma}^- + \h{\Gamma}^+ \left[ i V + \partial_t + \frac{e}{4m} F \right] + \h{\Gamma}^a \mathcal{D}_a \, ,  
\ee
where  $\mathcal{D}_a = \nabla_a - i e A_a$ is the $U(1)$ covariant spinor derivative on $\M$. We also define the Dirac operator on $\M$ with A flux as $\mathcal{D} = \Gamma^a \mathcal{D}_a$. 

According to the gamma matrices representation \eqref{eq:gamma_matrices_representation} we write a spinor $\h{\psi}$ on $\oM$ as 
\be \label{eq:oM_spinor_split}
\h{\psi} = \left( \begin{array}{c} \chi_1 \\ \chi_2 \end{array} \right) \, , 
\ee 
where  $\chi_1$ and $\chi_2$ are spinors on $\M$. Then the massless Dirac equation on $\oM$, $\h{D} \h{\psi} = 0$, can be written as 
\be 
\left( \begin{array}{cc} \mathcal{D} & \;\; \mathcal{O} + \partial_t \\ 
                                     im & - \mathcal{D} 
         \end{array} \right)  
\left( \begin{array}{c} \chi_1 \\ \chi_2 \end{array} \right) = 0 \, , 
\ee 
where the operator $\mathcal{O}$ is given by $\mathcal{O} = i V  + \frac{e}{4m} F$, and it reduces to two equations on $\M$: 
\be \label{eq:Levy-Lebond} \left\{ \begin{array}{lcl} 
\partial_t \chi_2 + \mathcal{O} \chi_2 + \mathcal{D} \chi_1  &=& 0 \, ,  \\  i m \chi_1 - \mathcal{D} \chi_2 &=&0  \, . \end{array} \right.  
\ee
This is the non-relativistic L\'evy-Leblond equation for a particle of mass $\tilde{m} = \frac{m}{2}$, in curved space, with scalar and vector potential  and with an additional term $\frac{e}{4m} F$. This extra term induces an anomalous gyromagnetic factor $g=3/2$ and was not present in the original work by L\'evy-Leblond. To see the value of the gyromagnetic factor one can in the second equation in eq.\eqref{eq:Levy-Lebond} find $\chi_1$ as function of $\chi_2$ and substitute it back in the first equation, thus getting 
\be 
\left[ E - \frac{1}{2\tilde{m}}(\Pi^\mu - e A^\mu)^2  - V - i \frac{3e}{8\tilde{m}} F - \frac{R}{8\tilde{m}} \right] \chi_2 = 0 \, , 
\ee 
where $E = i \partial_t$, $\Pi_\mu = - i \nabla_\mu$ is the momentum and $R$ the scalar curvature of $\M$. 

In \cite{DuvalHorvathyPalla1996} the anomalous gyromagnetic factor term does not appear, however there the setting is different: dimensional reduction is done starting from a massless Dirac equation with vector flux on $\h{M}$, while in the present case the vector potential originates from the metric and not directly from the massless Dirac equation. Therefore the massless Dirac equation considered in \cite{DuvalHorvathyPalla1996}  that is dimensionally reduced  is not the same as the one considered here. Situations where a dimensional reduction induces an anomalous gyromagnetic factor are not unknown, see \cite{GaryWiltshire1985} for an example and further references. 
 
This dimensional reduction gives a geometrical derivation of the L\'evy-Leblond equation. One of the reasons why this is ultimately possible is the fact that the Bargmann group -- the central extension of the Galilei group that leaves invariant the Schr\"{o}dinger equation and the L\'evy-Leblond equation -- can be embedded in the de Sitter  group $O(1,\dm +1)$ \cite{DuvalBurdetKunzlePerrin1985}.

\subsubsection{Relativistic Dirac equation: lift and reduction} 
The $\partial_t$ term, as seen in the previous section, is of main importance in order to obtain non-relativistic equations. In this section we seek to understand those cases when the massless Dirac equation on $\h{\M}$ can be dimensionally reduced to the, still relativistic, Dirac equation with flux on $\M$. As we will see this cannot always be done, differently from the non-relativistic case, the details of the reduction depending on the explicit form on the scalar and vector potentials. However, at least in the case of $V=m$ and $F = 0$ the dimensional reduction and its inverse, the lift, can always be performed. 
 
We start then by asking the following two conditions for $\h{\psi}$: 
\be \label{eq:spinor_partial_v_m} 
\partial_v \h{\psi} = im \h{\psi} \, , 
\ee 
\be 
\partial_t \h{\psi} = 0 \, ,  
\ee
which, as seen in the previous section, are compatible with the Dirac equation on $\h{\M}$. 

The Dirac operator reduces to 
\be 
\h{D} = i m  \h{\Gamma}^- + \h{\Gamma}^+ \left[ i V  + \frac{e}{4m} F \right] + \h{\Gamma}^a \mathcal{D}_a \, .   
\ee
Then the massless Dirac equation on $\oM$, $\h{D} \h{\psi} = 0$, can be written as 
\be \label{eq:Dirac_system_1} 
\left( \begin{array}{cc} \mathcal{D} & \;\; \mathcal{O} \\ 
                                     im & - \mathcal{D} 
         \end{array} \right)  
\left( \begin{array}{c} \chi_1 \\ \chi_2 \end{array} \right) = 0 \, , 
\ee 
where the operator $\mathcal{O}$ is given by $\mathcal{O} = i V  + \frac{e}{4m} F$. The general solution of $\h{D} \h{\psi} = 0$ is given by 
\be \label{eq:Dirac_general_solution}
\h{\psi} = \left( \begin{array}{c} - \frac{i}{m} \mathcal{D} \chi_2 \\ \chi_2 \end{array} \right) \, , 
\ee
where $\chi_2$ satisfies 
\be \label{eq:chi2_integrability}
\frac{i}{m} \mathcal{D}^2 \chi_2 = \mathcal{O} \chi_2 \, , 
\ee 
which is an integrability condition of \eqref{eq:Dirac_system_1}. 

In order to make contact with the Dirac equation with flux on $\M$ we impose the following condition on the generic spinor $\h{\psi}$ of \eqref{eq:oM_spinor_split} : 
\be \label{eq:spinor_projection}
i V \chi_1 = \mathcal{O} \chi_2 \, . 
\ee 
This allows to solve for $\chi_1$ if $\chi_2$ is known, and vice-versa since $\mathcal{O}$ is invertible for generic values of $F$, $V$ and $E$. The reason to ask for this condition is that, when it is satisfied, then the spinor $\chi_1$ satisfies 
\be 
(\mathcal{D} + i V) \chi_1 = 0 \, , 
\ee 
which is the Dirac equation with $V$ and $A$ flux on the base. It is worth noticing that the condition \eqref{eq:spinor_projection} is non-trivial and will be satisfied only for specific combinations of $V$ and $F$. 

To gain insight into the condition we can rewrite eq.\eqref{eq:spinor_projection} as a projection $P \h{\psi} = \h{\psi}$, where $P$ is the projector 
\be 
P = \frac{\mathbb{I} - i \, \frac{\mathcal{O}}{V} \h{\Gamma}^+ + i \, \frac{V}{\mathcal{O}} \h{\Gamma}^-}{2} \, . 
\ee 
$P$ is well defined when $- i V^{-1} \mathcal{O} = \mathbb{I} - i \frac{e}{4m} V^{-1} F$ is invertible, in which case it satisfies $P^2 = P$.

The following three alternative cases can occur. First, if  both $V=0$ and $F = 0 $ then eq.\eqref{eq:spinor_projection}  is always satisfied and $\chi_1$ satisfies the massless Dirac equation on the base with no flux, $D \chi_1 = 0$. Second, if instead $V \neq 0$ but $F = 0$ then $- i V^{-1} \mathcal{O} = \mathbb{I}$ and $P = \frac{1}{2}\left(\mathbb{I} + \Gamma^+ + \Gamma^-  \right)$. Then $P \h{\psi} = \h{\psi}$ has the symmetric solution $\chi_1 = \chi_2$. Third, if both $V\neq 0$ and $F\neq 0$ then for  generic values of $V$ and $F$ the operator $P$ will be well defined and the solution of $P \h{\psi} = \h{\psi}$ will be a twist of the symmetric case parameterised by $F$. 

In the symmetric case $V\neq 0$ and $F=0$ the projector $P$ satisfies the equation 
\be \label{eq:projector_useful}
\h{D} P = - \frac{\Gamma^+ + \Gamma^-}{2} \h{D} + i(m + V) \, , 
\ee 
which means that for generic values of $V$ it will not be possible to ask that for any spinor $\h{\psi}$ that satisfies the massless Dirac equation on $\h{\M}$ then its projection $P\h{\psi}$ will also satisfy the equation. The only exception to this is for $V = -m$ which corresponds to considering the Dirac equation on $\M$ for a particle of mass $m$. In this case if $\h{D}\h{\psi}=0$ then $\h{D} P \h{\psi}=0$ and from $P\h{\psi}$ we can construct a solution of the Dirac equation with mass on $\M$. Considering all possible such $\h{\psi}$ we can construct all the independent solutions on $\M$. Vice-versa given a solution of the Dirac equation with mass on $\M$ this can be lifted to a solution of the massless Dirac equation on $\h{\M}$ satisfying $P \h{\psi} = \h{\psi}$. On the other hand, a direct substitution of the condition $\chi_1 = \chi_2$ into the Dirac equation \eqref{eq:Dirac_system_1}  shows that the case $V = - m$ is the only one consistent with the spinor equations. 
It is interesting noticing that solutions of the equation $P\h{\psi} = 0$, that are orthogonal to the previous ones, satisfy the condition $ - i V \chi_1 = \mathcal{O} \chi_2$. Then these are associated to solutions of the Dirac equation on the base with $-V$ flux.

In the generic case $V\neq 0$ and $F\neq 0$ equation \eqref{eq:projector_useful} will also receive contributions proportional to $V^{-1}F$ and its derivatives. This opens the possibility of non-trivial solutions although these are not easy to analyse in the general case. One thing that is possible to do is to analyse the compatibility condition between the projection and the Dirac equation. When both hold then $\chi_2$ also has to satisfy the non-trivial equation 
\be 
 \frac{V}{m} \mathcal{D} \chi_2 = \mathcal{O} \chi_2 \, . 
\ee 
In section \ref{sec:examples} we show a simple example with non-zero $V$ and $F$ where the projection is compatible with the Dirac equation. 

We conclude the discussion showing how to lift a solution of the Dirac equation with flux to a solution of the massless Dirac equation on $\h{\M}$. Suppose there exists a given spinor $\chi_1$ on $\M$ satisfying the Dirac equation with flux 
\be 
\mathcal{D} \chi_1 + i V \chi_1 = 0 \, . 
\ee 
In order to upgrade $\chi_1$ to a spinor on $\oM$ we can define a spinor $\chi_2$ on $\M$  solving 
\be \label{eq:chi2_as_function_of_chi1}
\mathcal{D} \chi_2 = i m \chi_1 \, . 
\ee
Now in order for the spinor $\h{\psi} = (\chi_1, \chi_2)$ to satisfy the massless Dirac equation on $\h{\M}$ we need to ask the further condition \eqref{eq:spinor_projection}. Again the number of solutions will not be maximal due to compatibility conditions.

\subsection{Hidden symmetry operators} 
In this section we examine hidden symmetry operators of the two theories from the two complementary points of view of dimensional reduction and lift. From the former,  it is to be expected that not all hidden symmetries of the dynamics on $\h{\M}$ can be reduced to symmetries of the dynamics on $\M$ as the embedding of the lower dimensional theory in the higher dimensional one is a proper inclusion in terms of dynamics. For example there are KY and CCKY forms on $\h{\M}$ that arise as lifts of  KY and CCKY forms on $\M$ and yet such that their corresponding symmetry operators cannot be dimensionally reduced. This is consistent with the observation made in secs.\ref{sec:ansatz1} and \ref{sec:ansatz4} that  according to whether the form is even or odd different conditions are required in order for it to generate a hidden symmetry operator for the Dirac equation with flux on $\M$. Those symmetry operators that cannot be dimensionally reduced are built from even KY or odd CCKY tensors and are those that in lower dimension would generate anomalous symmetry operators. 
From the point of view of the lift of hidden symmetry operators instead we have seen in section \ref{sec:lift} that it is possible to have symmetry operators on $\M$ generated by CKY tensors that cannot be lifted to $\h{\M}$. 

At the end of the section we will discuss a simple example in which a CKY tensor on $\M$ that is neither strictly KY nor CCKY, when it exists, gives rise to a symmetry operator of the Dirac equation on $\M$ but this symmetry cannot be lifted to a symmetry operator of the Dirac equation on $\h{\M}$. While this is not a proof that such a lift is not possible we hope that such an example can help create an intuition on the underlying reason why the lift cannot be done. A proof follows from the results of section \ref{sec:Eisenhart} where it is shown no CKY tensor on $\h{\M}$ can be written in terms of a CKY form on $\M$ that is neither KY or CCKY. 

We start analysing dimensional reduction. We consider the KY form $\h{f}_1$ on $\h{\M}$ that is obtained by lifting a KY $p$--form $f$ on $\M$, as in sec.\ref{sec:ansatz1}, and its associated symmetry operator of the Dirac equation given by eq.\eqref{SOProp1}. From the results of that section, and using eqs.\eqref{eq:df},\eqref{eq:delta_f}, one can see that $\h{\delta} \h{f}_1 = 0$, and $\h{d} \h{f}_1 = df$. Then in terms of explicit components we can write 
\ba 
\h{S}_{\h{f}_1} &=& \frac{1}{(p-1)!} \left[ \h{\Gamma}^{B_1 \dots B_{p-1}} \h{f}_1^A {}_{B_1 \dots B_{p-1}} \h{\nabla}_A \right. \nn \\ 
&& \left. + \frac{1}{2(p+1)^2} \h{\Gamma}^{A_1 \dots A_{p+1}} \h{d}\h{f}_{1 \, A_1 \dots A_{p+1}} \right] \nn \\ 
& = &  \frac{(\sigma_3)^{p-1}}{(p-1)!} \otimes \left[ \Gamma^{b_1 \dots b_{p-1}} f^a {}_{b_1 \dots b_{p-1}} (\nabla_a - \frac{e}{m} A_a \partial_v ) \right. \nn \\ 
&& \left.+ \frac{1}{2(p+1)^2} \Gamma^{a_1 \dots a_{p+1}} df_{a_1 \dots a_{p+1}} \right] \, . 
\ea 
Notice that there is a term proportional to $(X^a \hook f) \h{\Gamma}^+ (X_a \hook F)$ coming from $\h{\nabla}$ that drops out due to eqs.\eqref{eq:ansatz1_useful1},\eqref{eq:ansatz1_useful2} and the identities \eqref{usefulProduct1}, \eqref{usefulProduct2}. The operator above is a symmetry operator of the Dirac operator on $\h{\M}$, in other words it transforms solutions of the equation in other solutions. However, in order to see whether it can be dimensionally reduced to a symmetry operator of the Dirac equation with flux on $\M$ we need to check whether its action commutes with the projection \eqref{eq:spinor_projection}. If it does, then its action generates an action on the space of solutions of the Dirac equation with flux into itself. 

First of all, we notice that when the higher dimensional spinor $\h{\psi}$ satisfies condition \eqref{eq:spinor_partial_v_m} then the action of $\h{S}_{\h{f}_1}$ is the same as that of 
\be 
(\sigma_3)^{p-1} \otimes \mathcal{S}_f \, , 
\ee 
where 
\ba \label{eq:symmetry_op_M} 
\mathcal{S}_f &=& \frac{1}{(p-1)!} \otimes \left[ \Gamma^{b_1 \dots b_{p-1}} f^a {}_{b_1 \dots b_{p-1}} \mathcal{D}_a \right. \nn \\ 
&& \left.  + \frac{1}{2(p+1)^2} \Gamma^{a_1 \dots a_{p+1}} df_{a_1 \dots a_{p+1}} \right] 
\ea 
 is an operator on $\M$ that formally is the same as a symmetry operator generated from $f$ but such that it uses the $U(1)$ covariant derivative instead of the spinor derivative. Let us then define a new solution $\h{\psi}^\prime$ by 
\be 
\h{\psi}^\prime = \left( \begin{array}{c} \chi_1^\prime \\ \chi_2^\prime \end{array} \right) = 
(\sigma_3)^{p-1} \otimes \mathcal{S}_f \left( \begin{array}{c} \chi_1 \\ \chi_2 \end{array} \right) = 
  \left( \begin{array}{c} \mathcal{S}_f \, \chi_1 \\ (-1)^{p-1} \mathcal{S}_f \, \chi_2 \end{array} \right) \, ,  
\ee 
and check whether $\h{\psi}^\prime$ satisfies \eqref{eq:spinor_projection}. First of all notice that $\left[ iV, \mathcal{S}_f \right] \propto (X^a \hook f) \partial_a V = dV \hook f = 0$. Then $ i \, V \chi_1^\prime = \mathcal{S}_f \, iV \chi_1$. 
Second, since $\h{\psi}$ satisfies the Dirac equation then 
\ba 
\mathcal{O} \chi_2^\prime &=& \frac{i}{m} \mathcal{D}^2 \chi_2^\prime = \frac{i}{m} \mathcal{D}^2 \left( (-1)^{p-1} \mathcal{S}_f \, \chi_2 \right) \nn \\ 
&& \hspace{-0.75cm} =  (-1)^{p-1} \, \frac{i}{m} \, \mathcal{S}_f \, \mathcal{D}^2 \, \chi_2 = (-1)^{p-1} \mathcal{S}_f \, \mathcal{O} \chi_2 \, . 
\ea 
In the last equality we have used the result found in \cite{DavidClaudePavel2011} that $\mathcal{S}_f$ graded commutes with $\mathcal{D}$, and therefore commutes with $\mathcal{D}^2$. Then 
\be 
 i \, V \chi_1^\prime = \mathcal{S}_f \, iV \chi_1 = \mathcal{S}_f \mathcal{O} \chi_2 = (-1)^{p-1} \mathcal{S}_f \, \mathcal{O} \chi_2^\prime \, . 
\ee 
This proves that the symmetry operator $\h{S}_{\h{f}_1}$ commutes with the projection \eqref{eq:spinor_projection} if and only if $p$ is odd. Then in this case it generates a symmetry operator of the Dirac equation with flux on $\M$, by looking at its action on the spinor $\chi_1$. Such action  is given in terms of the operator $\mathcal{S}_f$, which is the symmetry operator found in \cite{DavidClaudePavel2011}. In this reference it was found that a subset of the conditions found in sec.\ref{sec:ansatz1}, namely eqs.\eqref{eq:ansatz1_useful_dV}, \eqref{eq:ansatz1_KY}, \eqref{eq:ansatz1_useful2}, guarantee that $\mathcal{S}_f$ is a symmetry operator of the Dirac equation with flux, and the results of this section provide a geometrical alternative proof of the original result\footnote{The conditions found in \cite{DavidClaudePavel2011} are more general then those found here, the most general possible. Correspondingly, not all of the generalised CKY tensors found there can be lifted in the present context to CKY tensors on $\h{\M}$, as discussed in section \ref{sec:Eisenhart}.}. 
 
A similar analysis can be done with respect to the CCKY tensor $\h{f}_4$ discussed in sec.\ref{sec:ansatz4}. We consider the symmetry operator \eqref{eq:symmetry_operator_CCKY}. This can be written down explicitly and dimensionally reduced when $p$ is even. Now the conditions found in sec.\ref{sec:ansatz4} are required to see that the $F$ term present in the spinor derivative $\h{\nabla}$ drops out, and that the action of the operator commutes with that of the projection \eqref{eq:spinor_projection}. Again, the conditions found are a subset of those found in \cite{DavidClaudePavel2011} for the case of even tensors. 

With this we have shown concretely how not all the higher dimensional symmetry operators corresponding to CKY tensors can be dimensionally reduced to give symmetry operators in lower dimension. In particular this happens also for a subset of the tensors found in secs. \ref{sec:ansatz1}, \ref{sec:ansatz4}, which were obtained as a lift of CKY tensors on $\M$. Those lifted tensors such that their higher dimensional symmetry operators do not commute with the projection \eqref{eq:spinor_projection} are exactly those for which in lower dimension the corresponding symmetry operators are anomalous due to the presence of flux. This provides a geometrical interpretation of the anomaly. It should be noticed that the hidden symmetries compatible with the lift, given by odd KY  and even CCKY tensors, are exactly those which give rise to operators that fully commute with the Dirac operator \cite{MDP2011_1}. 

Now we turn to the last objective of this section, providing an intuitive explanation why not all of the symmetry operators found in lower dimension can be lifted. In order for the calculations to be simpler, suppose there is no flux: $V \equiv 0$, $F \equiv 0$. Let's assume that on $\M$ there exists a CKY $p$--form $\omega$ that is neither KY nor CCKY. Then from eq.\eqref{eq:CKY_graded_commutator} we know that 
 \be 
 DS_\omega = (-1)^{p-1} S_\omega D - \frac{(-1)^{p-1}}{n- p + 1} \delta\omega  D\, . 
\ee
Suppose we try to lift $S_\omega$ to an operator on $\h{\M}$ of the kind $\h{S} = (\sigma_3)^{p-1} \otimes S_w$. This is not the only possible way of doing the lift and therefore this is not a proof, but rather an illustration. Then 
\ba 
\h{D} \h{S} &=& \left( \h{\Gamma}^- \partial_v + \h{\Gamma}^+  \partial_t   + \sigma_3 \otimes D \right) (\sigma_3)^{p-1} \otimes S_w \nn \\ 
&=& (-1)^{p-1} \h{S} \h{D} + \frac{(-1)^{p}}{n- p + 1} (\sigma_3)^p \delta \omega D \, . 
\ea 
If $\h{\psi}$ is a generic solution of the higher dimension Dirac equation the first term is zero, but in general the second will not be since $\delta \omega \neq 0$ and $D \chi_2 \neq 0$. This corresponds to the fact that there is no CKY tensor on $\h{\M}$ that can be written in terms of such a form $\omega$ on $\M$. So the impossibility to perform the lift seems  related to the doubling of the spinor degrees of freedom in higher dimension and the fact that the higher dimensional Dirac equation mixes such degrees of freedom.


\section{Examples\label{sec:examples}}
In this section we present some examples of metrics with and without flux.

\subsection{The case $V=m$, $F=0$} 
When $V=m$, $F=0$ it is possible to use solutions of the massless Dirac equation on $\h{M}$ to build solutions of the Dirac equation with mass on $\M$. It is also possible to perform the inverse operation and lift solutions from $\M$ to $\h{M}$. There is a number of known examples of non-trivial spacetimes with Euclidean signature that admit CKY tensors. Following \cite{Semmelmann2002} we can take $\M$ to be: the sphere $S^\dm$, which admits both KY and CCKY forms; Sasakian manifolds, which admit a rank 2 CCKY form, as for example $S^1$--bundles over K\"{a}hler manifolds; K\"{a}hler and nearly K\"{a}hler spaces, which admit a rank 2 CKY form; $G_2$ and weak-$G_2$ manifolds; there is also a classification of compact manifolds of Riemannian signature that admit special KY forms (for the definition of special KY forms we refer the reader to \cite{Semmelmann2002}), these are Sasakian, nearly K\"{a}hler and weak--$G_2$ manifolds. We can also consider $\M$ to be a Riemannian space admitting Killing spinors, along the lines of \cite{Marco2004}, as Killing spinors generate a tower of KY and CCKY forms. In particular we can consider: all spaces of special holonomy (Calabi-Yau, hyperK\"{a}hler, $G_2$, $Spin(7)$); maximally symmetric spaces; compact spaces with positive curvature the cone over which is irreducible, which include the already mentioned 3-Sasaki manifolds when $\dm = 4m -1$, $m$ an integer, with hyperK\"{a}hler cone, Sasaki-Einstein manifolds when $\dm = 4m \pm 1$ with Calabi-Yau cone, the already mentioned almost K\"{a}hler case when $\dm = 6$ which has $G_2$ cone, and weak--$G_2$ which has $Spin(7)$ cone; there are also non-compact spaces with negative curvature, in this case either the hyperbolic space $H^\dm$ or a warped product $M = N \times \mathbb{R}$ with metric $ds^2 = e^{\mu y} ds_N^2 + dy^2$, where $\mu \in \mathbb{R} \ {0}$ and N is a complete, connected spin manifold which admits non trivial parallel spinors. 
 
Another non-trivial example of Riemannian metric with CKY forms is to consider $\M$ to be a manifold with the canonical metric, which as shown in \cite{HouriOotaYasui_uniqueness,DavidPavelFrolov2008_uniqueness} is the most general metric which admits a PCKY form, that is a rank 2 non-degenerate CCKY form. This metric depends on a set of one-variable functions, and when these are chosen so that the metric satisfies the Einstein vacuum equations with cosmological constant then one obtains the (Wick rotated) Kerr-NUT-(A)dS metric of  \cite{ChenEtal:2006cqg}. The canonical metric in $\dm = 2N + \epsilon$ dimensions, where $\epsilon = 0,1$, admits $N+ \epsilon$ Killing vectors and $N$ CCKY forms of even rank. By the results of section \ref{sec:Eisenhart} it is possible to lift all of these tensors to KY and CCKY tensors on $\h{M}$. The metric $\h{g}$ on $\h{\M}$ also has the two Killing vectors $\partial_v$, $\partial_t$. 
The Dirac equation on $\M$ admits separation of variables due to a complete set of mutually commuting operators one of which is the Dirac operator \cite{OotaYasui2007,MDP2011_2}. Then if $\chi_1$ is a solution of the massive Dirac equation on $\M$ one can upgrade it to a solution of the massless Dirac equation on $\h{M}$ by setting $\chi_2 = \chi_1$.

\subsection{The case $V$ a central potential, $F=0$\label{sec:example_central}} 
Consider a flat Riemannian space $\M = \mathbb{R}^\dm$. Suppose the electromagnetic field is zero, $F=0$ and the potential $V$ is central, $V = V(r)$, with $r^2 = x_\mu x^\mu$. Then we can consider the rank $\dm -1$ KY tensor 
\be 
f_{\lambda_1 \dots \lambda_{n-1}} = x^\mu \, \epsilon_{\mu \lambda_1 \dots \lambda_{n-1}} \,  \, , 
\ee 
built using the totally antisymmetric Levi-Civita $\dm$--form $\epsilon_{\lambda_1 \dots \lambda_n}$. There is a conserved tensor associated to any geodesic on $\M$, which is given by $C_{\lambda_1 \lambda_{n-2}} =  \dot{x}^\mu f_{\mu \lambda_1 \dots \lambda_{n-2} }$. When $\dm = 3$ the conserved quantity is proportional to the angular momentum and therefore this example includes cases such as the Kepler problem or the harmonic oscillator; for $d>3$ we can think of this as a generalisation of angular momentum. That the quantity is conserved can be seen by direct differentiation, noticing that the geodesic equation implies that $\ddot{x}$ is proportional to $\vec{\nabla} V = \frac{dV}{dr} \frac{\vec{x}}{r}$, and therefore 
\be 
\dot{C}_{\lambda_1 \dots \lambda_{n-2}} \propto \frac{dV}{dr}\frac{x^\mu}{r} x^\nu  \epsilon_{\mu \nu \lambda_1 \dots \lambda_{n-2}} \,  = 0 \, . 
\ee 
This condition is equivalent to $dV^\sharp \hook f = 0$, which is the same as equation \eqref{eq:ansatz1_useful_dV} and guarantees that we can promote $f$ to a KY tensor on the Eisenhart-Duval lift manifold $\h{M}$, and build conserved tensors for the null geodesic motion on $\h{M}$.

\subsection{The case $V\neq 0$, $F\neq 0$} 
We present here a very simple example with the intent of showing that the projection \eqref{eq:spinor_projection} does indeed admit non-trivial solutions. From the construction of this example however it will become clear that interesting non-trivial solutions will in general require more effort. 
 
We take $\M$ to be three dimensional flat Euclidean space. The Gamma matrices are $\Gamma_1 = \sigma_1$, $\Gamma_2 = \sigma_2$, $\Gamma_3 = \sigma_3$, the Pauli matrices. We consider a magnetic field $F = \phi(x) \sigma_1 \sigma_2 = i \phi \, \sigma_3$, and as an ansatz for the spinor $\chi_2$ we take the spinor $\chi_2 = (1,0)$. The projection becomes 
\be 
i V \chi_1 = i (V + \frac{e\phi}{4m} ) \chi_2 = i V \tilde{\phi} \chi_2  \, , 
\ee 
where we have defined the function 
\be 
\tilde{\phi} = 1+ \frac{e\phi}{4mV} \, . 
\ee 
The Dirac equation on $\h{M}$ gives 
\ba 
&& D\chi_1 + i V \chi_1 = 0 \, , \nn \\ 
&& D \chi_2 = i m \chi_1 \, . 
\ea 
One can check by direct substitution that these two coupled equations are compatible with the projection only if 
\be 
\left( i V + d(\ln \tilde{\phi}) \right) \chi_2 = - im \tilde{\phi} \chi_2 \, .  
\ee 
Then it must be $\tilde{\phi} = \tilde{\phi}(x_3)$ in which case the equation above becomes an equation for the function $\tilde{\phi}$: 
\be 
 i V + d(\ln \tilde{\phi})  = - im \tilde{\phi}  \, .  
\ee 
But $V$ and $\tilde{\phi}$ are both real, and therefore it must be that $\tilde{\phi}$ is constant and that $V = - m \tilde{\phi}$. Then also $V$ and the magnetic field are constant. The Dirac equation on $\h{M}$ reduces to the single equation 
\be 
D \chi_2 = i m \tilde{\phi} \chi_2 \, , 
\ee 
where $m \tilde{\phi}$ plays the role of an effective mass and a solution is given by 
\be 
\left( \begin{array}{c} e^{im\tilde{\phi} x_3} \\ 0 
\end{array} \right) \, . 
\ee


\section{Discussion and conclusions\label{sec:conclusions}}
In this paper we have studied the Eisenhart-Duval lift from the point of view of hidden symmetries of the Dirac equation and have gained insight on the relationship between the procedure of lift/oxidation, its inverse procedure of reduction, and the symmetry operators of the Dirac equation with flux. 
 
We have shown how the massless Dirac equation on $\h{M}$ can be dimensionally reduced to a non-relativistic L\'evy-Leblond equation on $\M$. We have also discussed those cases where it is possible to obtain on $\M$ again a Dirac equation. When $V = -m$ and $F=0$ it is always possible to obtain by reduction the massive Dirac equation on $\M$, and to lift the hidden symmetries on $\M$ to hidden symmetries on $\h{\M}$. However, for generic fluxes the Dirac equations in lower and higher dimension are not equivalent systems when considered from the point of view of their dynamics in phase space, since each theory can have hidden symmetries that are not present in the other. We have shown an example where the impossibility to lift some CKY tensors in the absence of flux is  related to the doubling of the spinor degrees of freedom in higher dimension and the way that the higher dimensional Dirac equation mixes such degrees of freedom, and examples where the impossibility to dimensionally reduce instead is related to the fact that symmetries of the Dirac equation with flux on $\M$ present anomalies and not all KY and CCKY tensors are allowed. Whenever it is possible to either lift or dimensionally reduce the symmetry operators then we find a geometrical relation between the symmetry operators of the two theories, one with flux and the other without. The situation is different from what happens analysing  the Eisenhart-Duval lift for a scalar particle: in that case all hidden symmetries of the lower dimensional theory can be lifted to hidden symmetries in higher dimension. 
 
A by-product of this analysis is that we can build new Lorentzian metrics with KY and CCKY tensors, by lifting KY and CCKY tensors defined on Riemannian metrics. We also have presented a classification of the most general CKY tensor for the Eisenhart-Duval metric in terms of a set of equations for forms on the base manifold, both in the $v$, $t$ independent case and in the general one. 

There is a number of questions left open in this work. One of these is: what is the actual form of generic solutions of the CKY equations on $\h{\M}$ in terms of forms on $\M$, and do these solutions give any non-trivial generalisation of the CKY equations? Another one is related to supergravity solutions: it is known how spacetimes with a null covariantly constant (Killing) vector can provide supersymmetric solutions of supergravity theories. It would be interesting to know if the present construction can be used in the context of supergravity to either provide new solutions or discuss existing ones in terms of hidden symmetries. Also it would be interesting to know if the tools used in the present analysis can also be used to study the more general class of Kundt spacetimes, see \cite{BrannlundColeyHervik200} for a recent discussion of their role in supergravity and string/M--theory.

\vspace*{1ex}
\section*{Acknowledgments}
The author would like to thank G. W. Gibbons for pointing out to him some of the open questions related to the Eisenhart-Duval lift and CKY tensors; P. Horv\'athy for numerous suggestions; D. Kubiz\v{n}\'ak and C. M. Warnick for reading the manuscript and useful discussions. Part of this work has been done during the author's stay at ICTP-SAIFR in S\~ao Paulo, hospitality there and financial support are kindly acknowledged. The author is partially funded by Fapemig under the project CEX APQ 2324-11.


\section*{Appendices}
\appendix

\section{Eisenhart-Duval metric\label{apdx:Eisenhart_metric}}
The non-zero Christoffel symbols for the Eisenhart-Duval lift metric \eqref{eq:Eisenhart_metric} are 
\be 
\begin{split}
& \oGamma^v_{tt} = - \frac{e}{m^2} A_\lambda \partial^\lambda V \, , \\ 
& \oGamma^v_{\mu t} = \frac{1}{2} \left( \frac{e^2}{m^2} A_\lambda F^\lambda_{\;\; \mu} - \frac{2}{m} \partial_\mu V \right) \, , \\ 
& \oGamma^v_{\mu \nu} = \frac{e}{m} \nabla_{(\mu} A_{\nu)}  \, , \\  
& \oGamma^\lambda_{tt} = \frac{1}{m} \partial^\lambda V \, , \\ 
& \oGamma^\lambda_{\mu t} = -\frac{e}{2m} F^\lambda_{\;\;\mu} \, , \\ 
& \oGamma^\lambda_{\mu\nu} = \Gamma^\lambda_{\mu\nu} \, . 
\end{split}
\ee
The covariantly constant Killing vector is $\partial_v$ and its associated 1--form is $(\partial_v)^\flat = dt$. 
A convenient choice for the vielbeins is 
\be \label{eq:Eisenhart_vielbeins}
\begin{split}
& \hat{e}^+ =  dt \, , \\ 
& \hat{e}^- = dv - \frac{V}{m} dt + \frac{e}{m} A_\mu dx^\mu \, , \\ 
& \hat{e}^a = e^a \, , 
\end{split}
\ee
where $\left\{e^a, a=1, \dots, n\right\}$ is a set of vielbeins for $\mathcal{M}$, and the $n+2$--dimensional Minkowski metric $\hat{\eta}_{AB}$ has the following non-zero entries: $\hat{\eta}_{+-}=\hat{\eta}_{-+}=1$, $\hat{\eta}_{ab} = \eta_{ab}$. The corresponding dual basis vectors are: 
\be
\begin{split}
& \left( \hat{e}^+  \right)^\sharp = \oX^+ =  \partial_v \, , \\ 
& \left( \hat{e}^- \right)^\sharp = \oX^- = \frac{V}{m} \partial_v + \partial_t \, , \\ 
& \left(\hat{e}^a \right)^\sharp = \oX^a =  - \frac{e}{m} A^a \partial_v + \left(e^a\right)^{\sharp g}   \, .  
\end{split}
\ee
These are related to the inverse vielbein $\oE^M_A$ by 
\be 
\oE_A = \hat{\eta}_{AB} \oX^B \, . 
\ee 

From eq.\eqref{eq:Eisenhart_vielbeins} we can read the non-zero coefficients of the spin--connection: 
\be \label{eq:spin_connection}
\begin{split}
& \oomega_{+a} = - \frac{1}{m} \partial_a V \, \hat{e}^+ + \frac{e}{2m} F_{ab} e^b \, , \\ 
& \oomega_{ab} = \omega_{ab} - \frac{e}{2m} F_{ab} \hat{e}^+ \, . 
\end{split} 
\ee


\section{Hodge duality\label{apdx:Hodge_duality}}
In this section we display identities for Hodge duality on $\M$, these are straightforwardly generalised to the Hodge duality on $\h{\M}$ with the appropriate change of signature and number of dimensions.  

The Levi-Civita tensor ${\eps}$  is an antisymmetric ${\dm}$-form satisfying
\begin{equation}\label{HD}
    \eps\cwedge{n}\eps = s\,\dm!\; , 
\end{equation}
where ${s}$ is the product of signs in the metric signature. With this the Hodge dual of a homogeneous ${p}$-form ${\omega}$ can be defined as 
\begin{equation}\label{HDdef}
    *\omega = \frac1{p!}\,\omega\cwedge{p}\eps\; . 
\end{equation}
It follows that
\ba\label{epsids}
    && **\omega = s (-1)^{p(\dm-p)}\omega\;,\quad
    *1 = \eps\;, \nn \\ 
&&  *\eps = s\;,\quad
    \eps^2 =  (-1)^{[\frac\dm2]}\, s\;.
\ea
The Hodge duality operation exchanges a wedge product into an interior product and vice-versa: for any vector $X^a$ dual to a vielbein $e^a$ it holds 
\ba\label{HDwedgehook}
   && * (e^a\wedge\omega) = (-1)^p X^a\hook (*\omega)\;, \nn \\
 &&   * (X^a\hook\omega) = (-1)^{p+1} e^a\wedge (* \omega)\;.
\ea
Also it transforms a contracted wedge product into a contracted wedge product. If  ${\alpha}$, ${\beta}$ are homogeneous forms of degrees ${p}$ and ${q}$, then 
\ba\label{HDcwedge}
    *\bigl(\alpha\cwedge{k}\beta)
     & = & \frac{k!}{(q-k)!}(-1)^{k(n-q)}(*\alpha)\cwedge{q{-}k}\beta \nn \\
      &=& \frac{k!}{(p-k)!}(-1)^{p(q-k)}\alpha\cwedge{p{-}k}(*\beta)\;.
\ea


\section{Differentiation of forms} 
Consider a $p$--form $\hat{f}$ on $\oM$ that only has components on $\M$, according to 
\ba \label{eq:fhat_equal_f} 
\hat{f} &=& f_{M_1 \dots M_p}(v,t,x^\mu) \hat{e}^{M_1} \wedge \dots \wedge \hat{e}^{M_1} \nn \\ 
&=& f_{\mu_1 \dots \mu_p} (v,t,x^\mu) e^{\mu_1} \wedge \dots \wedge e^{\mu_p} \, . 
\ea
When there is no $v$, $t$ dependence then this yields a form $f$ on $\M$. We can explicitly calculate the components of the tensor $\hat{\nabla} \hat{f}$. We start with 
\ba \label{eq:nabla_minus_f}
\onabla_- \hat{f}_{\mu_1 \dots \mu_p} &=& \oE_{-}^M \onabla_M \h{f}_{\mu_1 \dots \mu_p} = \oX^{+ M} \onabla_M \h{f}_{\mu_1 \dots \mu_p} \nn \\ 
&=&  \onabla_v \h{f}_{\mu_1 \dots \mu_p} = \partial_v \h{f}_{\mu_1 \dots \mu_p} + \left( \oGamma_v \cdot f\right)_{\mu_1 \dots \mu_p} \nn \\ 
&=& \partial_v \h{f}_{\mu_1 \dots \mu_p} \, .  
\ea
A similar calculation yields  
\be \label{eq:nabla_plus_f}
\h{\nabla}_+ \h{f} = \left(  \frac{V}{m} \partial_v + \partial_t \right) \h{f} +  \frac{e}{2m} \left(F \cwedge{1} \h{f} \right) - \frac{1}{m} \h{e}^+ \wedge ( dV^\sharp \hook \h{f} ) \, ,  
\ee 
and 
\be \label{eq:nabla_a_f}
\onabla_a \hat{f} = - \frac{e}{m} A_a \partial_v \h{f} +  \nabla_a \h{f} - \frac{e}{2m} \h{e}^+ \wedge [ (X_a \hook F) \hook \h{f}  ] \, . 
\ee
With this we are able to calculate 
\ba \label{eq:df}
\hat{d} \hat{f} = \hat{e}^A \wedge \onabla_A \hat{f} &=& e^+ \wedge  \left(   \frac{V}{m} \partial_v + \partial_t \right) \hat{f}  + e^- \wedge \partial_v \hat{f} \nn \\ 
&& +  \left( - \frac{e}{m} A \wedge \partial_v \hat{f} + d \hat{f} \right) \, , 
\ea
and 
\be \label{eq:delta_f}
\hat{\delta} \hat{f} = - \left( \hat{e}^A\right)^\sharp \hook \onabla_A \hat{f} = \frac{e}{m} A^\sharp \hook \partial_v \hat{f} + \delta \hat{f} + \frac{e}{2m} e^+ \wedge (F \cwedge{2} \h{f})  \, . 
\ee
Analogous relations for the forms $\hat{e}^+$ and $\hat{e}^-$ are 
\ba \label{eq:nabla_eforms}
\onabla_A \hat{e}^+ &=& 0 \qquad \forall M = +,-, a \, , \nn \\ 
\onabla_- e^- &=& 0 \, , \nn \\ 
\onabla_+ e^- &=& \frac{1}{m} dV \, , \nn \\ 
\onabla_a e^- &=& \frac{e}{2m} X_a \hook F \, .   
\ea


\section{The full CKY equation\label{apdx:GeneralCase}}

From the $\h{\nabla}_-$ equation we get four identities: 
\ba
&& (--) \; \partial_v \rho^- = 0 \, , \nn \\  
&& (-+) \; \frac{p}{p+1} \partial_v \rho^+ = \frac{1}{p+1} \left(  - \left( \frac{V}{m} \partial_v + \partial_t \right) \rho^- + \right.   \nn \\ 
&& \left. \frac{e}{m} A \wedge \partial_v g - dg \right) - \frac{1}{\dm +3 - p}  \left( \frac{e}{m} A^\sharp \hook \partial_v f + \delta f \right.\nn \\ 
&& \left. - \partial_v \rho^+ - \left( \frac{V}{m} \partial_v + \partial_t \right) \rho^- - \frac{e}{2m} F \cwedge{1} \rho^- \right)  \, , \nn \\ 
&& (-\pm) \; (\dm + 2 - p) \partial_v g =  \left( \delta \rho^-  + \frac{e}{m} A^\sharp \hook \partial_v \rho^- + \pa_v g \right)  \, , \nn \\ 
&& (-a) \; \frac{p}{p+1} \partial_v f =  \frac{1}{p+1} \left(  \frac{e}{m} A \wedge \pa_v \rho^- - d \rho^- \right)  \, . 
\ea 
We can think of these as equations for the partial $v$ derivatives of the forms $f$, $\rho^\pm$, $g$. 
Then from $\h{\nabla}_+$: 
\ba 
&& (+-) \;  \frac{p}{p+1} \left( \frac{V}{m} \partial_v + \partial_t \right) \rho^-  + \frac{e}{2m} F \cwedge{1} \rho^- = \frac{1}{p+1} \left( - \pa_v \rho^+  \right. \nn \\ 
&& \left.     - \frac{e}{m} A \wedge \pa_v g + dg \right)  - \frac{1}{\dm + 3 - p} \left( \delta f + \frac{e}{m} A \hook \pa_v f - \pa_v \rho^+ \right. \nn \\ 
&& \left. - \left( \frac{V}{m} \partial_v + \partial_t \right) \rho^- - \frac{e}{2m} F \cwedge{1} \rho^- \right) \, , \nn \\ 
&& (++) \; - \frac{dV^\sharp}{m}  \hook f + \left( \frac{V}{m} \partial_v + \partial_t \right) \rho^+ + \frac{e}{2m} F \cwedge{1} \rho^+  \nn \\ 
&& + \frac{dV}{m}  \wedge g = 0 \, , \nn 
\ea 
\ba
&& (+ \pm) \; \frac{\dm +2 - p}{\dm + 3 - p} \, \left[ \frac{dV^\sharp}{m} \hook \rho^- +  \left( \frac{V}{m} \partial_v + \partial_t \right) g \right] + \frac{e}{2m} F \cwedge{1} g  \nn \\ 
&&  =  - \frac{1}{\dm + 3 - p} \left( - \frac{e}{2m} F \cwedge{2} f + \frac{e}{m} A^\sharp \hook \pa_v \rho^+ + \delta \rho^+ \right) \nn \\ 
&& (+a) \; \frac{p}{p+1} \left( \frac{V}{m} \partial_v + \partial_t \right) f + \frac{e}{2m} F \cwedge{1} f + \frac{p}{p+1} \frac{dV}{m} \wedge \rho^-  \nn \\ 
&& = \frac{1}{p+1} \left(  \frac{e}{m} A \wedge \pa_v \rho^+ - d \rho^+  - \frac{e}{m} F \wedge g \right) \, . 
\ea 
These are equations for the partial $t$ derivatives. Lastly the $\h{\nabla}_a$ equation gives 
\ba 
&& (a-) \, - \frac{e}{m} A_a \pa_v \rho^- + \nabla_a \rho^- =    \frac{X_a}{p+1} \hook \left( - \pa_v f  \right. \nn \\ 
&& \left. - \frac{e}{m} A \wedge \pa_v \rho^- + d\rho^- \right)  - \frac{e_a}{\dm + 3 - p} \wedge \left( \delta \rho^- \right.     \nn \\ 
&& \left. + \frac{e}{m} A^\sharp \hook \pa_v \rho^- + \pa_v g \right) \, , \nn \\ 
&& (a+) \; - \frac{e}{2m} (X_a \hook F) \cwedge{1} f  + \nabla_a \rho^+ - \frac{e}{m} A_a \pa_v \rho^+ \nn \\ 
&&+ \frac{e}{2m} (X_a \hook F) \wedge g =    \frac{X_a}{p+1} \hook \left[  - \left( \frac{V}{m} \partial_v + \partial_t \right) f \nn \right. \\ 
&& \left. - \frac{e}{m} A \wedge \pa_v \rho^+ +  d \rho^+  - \frac{dV}{m} \wedge \rho^-  + \frac{e}{m} F \wedge g \right]  \nn \\ 
&& - \frac{e_a}{\dm + 3 - p} \wedge \left[  -\frac{e}{2m} F \cwedge{2} f  + \frac{e}{m} A^\sharp \hook \pa_v \rho^+  + \delta \rho^+ - \frac{dV^\sharp}{m} \hook \rho^-   \right] \, , \nn \\
&& (a \pm) \;  \nabla_a g + \frac{e}{2m} (X_a \hook F) \cwedge{1} \rho^-  - \frac{e}{m} A_a \pa_v g = \frac{X_a}{p+1} \hook \left( - \pa_v \rho^+ \right. \nn \\ 
&& \left. + \left( \frac{V}{m} \partial_v + \partial_t \right) \rho^-  - \frac{e}{m} A \wedge \pa_v g + dg \right) 
    - \frac{e_a}{\dm + 3 - p} \wedge \left( \delta g \right. \nn \\ 
&& \left. + \frac{e}{m} A^\sharp \hook \pa_v g + \frac{e}{2m} F \cwedge{2} \rho^-    \right) \, , \nn \\
&& (ab) \; - \frac{e}{m} A_a \pa_v f +  \nabla_a f + \frac{e}{2m} (X_a \hook F) \wedge \rho^- =   \frac{X_a}{p+1} \hook \nn \\ 
&& \left( - \frac{e}{m} A \wedge \pa_v f   + df + \frac{e}{m} F \wedge \rho^- \right)    - \frac{e_a}{\dm + 3 - p} \wedge \left( \delta f \right. \nn \\ 
&& \left. \frac{e}{m} A^\sharp \hook \pa_v f    - \pa_v \rho^+ - \left( \frac{V}{m} \partial_v + \partial_t \right) \rho^-  - \frac{e}{2m} F \cwedge{1} \rho^-  \right)   \, .
\ea 
Further simplifications can be obtained taking each of the four $\h{\nabla}_a$ equations, and calculating the product $e^a \wedge$ and $X^a \hook$, summing over $a$. This gives equations for the differential and co-differential of $f$, $\rho^\pm$, $g$ that can be put back in the other CKY equations.


\vspace*{-1ex}



\providecommand{\href}[2]{#2}\begingroup\raggedright

\end{document}